\documentclass[12pt]{article}
\pdfoutput=1
\usepackage{jheppub_X}

\usepackage[utf8]{inputenc}

\usepackage{adjustbox}
\usepackage[normalem]{ulem}
\usepackage{cleveref}
\crefname{table}{Table}{Tables}
\crefname{equation}{Eq.}{Eqs.}
\crefname{appendix}{App.}{Apps.}
\crefname{section}{Sec.}{Secs.}
\crefname{figure}{Fig.}{Figs.}

\usepackage{setspace}
\usepackage{tikz}
\usetikzlibrary{decorations.markings}
\usepackage{tcolorbox}
\usepackage{bm}
\usepackage{xfrac}
\usepackage{pbox}
\usepackage{graphicx}
\usepackage{bbold}
\usepackage{slashed,braket}
\usepackage{physics}
\usepackage{graphbox}

\allowdisplaybreaks

\makeatletter
\g@addto@macro\bfseries{\boldmath}
\makeatother

\definecolor{colorTC}{rgb}{.2,.7,.2}

\def\eg{\textit{e.g.}}
\def\ie{\textit{i.e.}}

\newcommand{\lag}{\ensuremath{\mathcal{L}}}

\newcommand{\s}{\hspace{0.8pt}}

\newcommand{\R}{R}
\newcommand{\Rhat}{\hat{R}}

\newcommand{\hg}[1]{{\dot{#1}}}
\newcommand{\HH}[2]{{\mathcal{H}_{\mathbf{c},#1}^{(#2)}}}

\preprint{CERN-TH-2023-058}

\title{
\huge
Effective Field Theory of the \\
Two Higgs Doublet Model
}

\author[a]{Ian Banta,}
\author[b,c,d]{Timothy Cohen,}
\author[e,a]{Nathaniel Craig,}
\author{\\[6pt] }
\author[f]{\hspace{-10pt} Xiaochuan Lu,}
\author{and}
\author[g]{Dave Sutherland \hspace{-4pt}}

\affiliation[a]{\fontsize{10}{10}\selectfont Department of Physics, University of California, Santa Barbara, CA 93106, USA}
\affiliation[b]{\fontsize{10}{10}\selectfont Theoretical Physics Department, CERN, 1211 Geneva, Switzerland}
\affiliation[c]{\fontsize{10}{10}\selectfont Theoretical Particle Physics Laboratory, EPFL, 1015 Lausanne, Switzerland}
\affiliation[d]{\fontsize{10}{10}\selectfont Institute for Fundamental Science, University of Oregon, Eugene, Oregon 97403, USA}
\affiliation[e]{\fontsize{10}{10}\selectfont Kavli Institute for Theoretical Physics, University of California, Santa Barbara, CA 93106, USA}
\affiliation[f]{\fontsize{10}{10}\selectfont Department of Physics, University of California, San Diego, La Jolla, CA 92093, USA}
\affiliation[g]{\fontsize{10}{10}\selectfont School of Physics and Astronomy, University of Glasgow, Glasgow G12 8QQ, UK}

\abstract{
We revisit the effective field theory of the two Higgs doublet model at tree level.
The introduction of a novel basis in the UV theory allows us to derive matching coefficients in the effective description that resum important contributions from the Higgs vacuum expectation value.
The new basis typically provides a significantly better approximation of the full theory prediction than the traditional approach that utilizes the Higgs basis, particularly for alignment away from the decoupling limit.
}

\begin{document}
\maketitle
\flushbottom
\setcounter{page}{2}

%%%%%%%%%%%%%%%%%%%%%%%%%%%%%%%%%%%%%%%%%%%%%%%%%%%%%%%%%%%%%%%%%%%%%%%%%%%%%%%%
\section{Introduction}
\label{sec:intro}
%%%%%%%%%%%%%%%%%%%%%%%%%%%%%%%%%%%%%%%%%%%%%%%%%%%%%%%%%%%%%%%%%%%%%%%%%%%%%%%%

They say good things come in pairs. This is certainly true in the search for new particles, where a second Higgs doublet has long been a quintessential candidate for physics beyond the Standard Model (BSM). The resulting two Higgs doublet model (2HDM) has been a subject of active study since its introduction in the 1970's (the original goal was to provide a model with spontaneous CP violation that could explain the CKM phase)~\cite{Lee:1973iz, Weinberg:1976hu}. Two Higgs doublet models arise in many motivated extensions of the Standard Model and provide perhaps the simplest realization of a spin-0 sector that matches the richness of the observed spin-1/2 and spin-1 sectors. Subsequent exploration of the many facets of 2HDMs has given rise to a vast literature; see \emph{e.g.}~\cite{Branco:2011iw} for a classic review.

The model predicts the addition of four new physical degrees of freedom to the Standard Model.  The existence of these BSM states may be inferred from both their direct production and their indirect imprints on the couplings of the already observed Higgs boson.  Over time, dedicated searches for these experimental signatures have been used to constrain the allowed parameter space.  This has engendered the generic expectation that the extra Higgs bosons in the 2HDM are likely to be at least several hundreds of GeV (barring a number of known loopholes in certain regions of parameter space).  If the new states in the 2HDM are heavy compared to the electroweak scale, an Effective Field Theory (EFT) description becomes a useful way to characterize the resulting deviations from the Standard Model at low energies.

Subtleties arise when matching a 2HDM onto an EFT with only one light Higgs boson. Integrating out the BSM Higgs bosons generically leads to an EFT for the observed Higgs boson $h$ in which electroweak symmetry is nonlinearly realized, often referred to as the Higgs EFT (HEFT). Alternately, integrating out an $SU(2)_L$ doublet of approximate mass eigenstates {\it can} lead to an EFT for a Higgs doublet $H$ in which electroweak symmetry is linearly realized, often referred to as the Standard Model EFT (SMEFT). In this case, the misalignment between the gauge and mass eigenstates is encoded by irrelevant operators in the EFT. Whenever SMEFT is admissible, it is often the preferred framework due to its compact parameterization and more transparent power-counting.

In a general 2HDM, there is a global $U(2)$ flavor symmetry acting on the two Higgs doublets.  Hence, there are infinitely many different basis choices one can specify in the UV description from which an infinite number of EFTs can be derived by integrating out one doublet.  These EFTs are only formally equivalent when the full tower of effective operators are included; different choices lead to different EFT Wilson coefficients and potentially different linearly realized symmetries. 

Given the freedom to choose a UV basis, what constitutes a good choice? Among many possible criteria, two stand out. First, the relative advantages of SMEFT over HEFT makes it preferable to choose a basis in which the low-energy theory is SMEFT, provided such a basis exists. Second, a good basis should allow the resulting EFT to accurately reproduce the effects of the full theory with as few operators as possible (\eg~at low orders in the EFT expansion).

In previous literature \cite{Gorbahn:2015gxa, Brehmer:2015rna, Egana-Ugrinovic:2015vgy, Belusca-Maito:2016dqe, Dawson:2017vgm, Dawson:2022cmu}, satisfying the first criterion has favored a particular basis for constructing 2HDM EFTs.  Integrating out a doublet that acquires a vacuum expectation value implies that the low-energy theory does not in general contain an electroweak symmetric point and thus requires HEFT instead of SMEFT. This fate can be avoided by using the {\it Higgs basis} \cite{Georgi:1978ri}, for which the light doublet contains all of the vacuum expectation value that breaks electroweak symmetry.\footnote{As emphasized in \cite{Egana-Ugrinovic:2015vgy}, this definition of the Higgs basis leaves a $U(1)_{\rm PQ}$ subgroup of the original $U(2)$ flavor symmetry intact, leading to a $U(1)$ family of Higgs bases.} Furthermore, the Higgs basis and the mass eigenstate basis become approximately aligned in the decoupling limit \cite{Haber:1989xc, Haber:1994mt} of CP-conserving 2HDMs, making the Higgs basis sensible for constructing the 2HDM SMEFT in this limit. However, exclusive use of the Higgs basis to meet our first criterion often makes it hard to meet the second criterion. The Higgs basis typically results in a poorly-convergent EFT expansion away from the decoupling limit even when SMEFT is formally appropriate for describing the low-energy theory. A more convergent EFT expansion can be obtained away from the decoupling limit by integrating out heavy mass eigenstates, but this generically yields HEFT. This tension has been a long-standing obstruction to the general EFT treatment of 2HDM.  

For better insight, it helps to recognize that the two criteria involve different points in field space. The origin in field space (where electroweak symmetry is restored) is essential for determining whether SMEFT can describe the low-energy theory, while our physical vacuum determines the composition of the mass eigenstates. It is therefore useful to rethink the basis choice in terms of a trajectory in field space that connects the origin, where electroweak symmetry is linearly realized, to the physical vacuum.  This motivates interpreting the field space of the theory in a geometric language where the EFT defines a submanifold of the UV description, as detailed in~\cite{Cohen:2020xca}.  The submanifold picture presents a new perspective on matching calculations: instead of integrating out approximate mass eigenstates or fields without vevs, one instead attempts to find a basis in the full theory that yields a simple parameterization of the EFT submanifold.

In this paper, we follow this strategy and identify a new basis for the 2HDM that simplifies integrating out the BSM states and matching to SMEFT (when possible) while also vastly improving convergence away from the decoupling limit. The key observation is that when there is a charge-preserving global minimum, there exists a basis choice for which (the zero-derivative part of) the classical solution of the heavy Higgs doublet is a \emph{linear} function of the light Higgs doublet; this defines what we call the ``straight-line'' (SL) basis.\footnote{While SL nominally denotes ``straight-line,'' four of the five authors would prefer to think of it as standing for ``SutherLand'', after its discoverer. The fifth author is too modest to contemplate naming a basis after himself. We leave it to the reader to decide.}  This basis --- which can be defined in {\it any} 2HDM with a charge-conserving global minimum --- unsurprisingly simplifies the matching calculation. 

Whether the EFT that results from matching in the SL basis can be SMEFT-like (linearly realizing electroweak symmetry) or must be HEFT-like depends on the parameters of the 2HDM itself; the SL basis is useful in either case. Since the vev of the heavy Higgs doublet vanishes at the same point as the vev of the light Higgs doublet (preserving an electroweak symmetric point in the EFT even though the heavy doublet acquires a vev elsewhere on the EFT submanifold), the SL basis satisfies our first criterion by enabling matching onto a SMEFT-like EFT whenever the parameters of the 2HDM admit it. This is not guaranteed in the Higgs basis, for which matching may lead to a HEFT-like EFT even if the 2HDM admits a SMEFT-like description. 

 As we will see, matching in the SL basis also satisfies our second criterion by resumming the zero-derivative Higgs field dependence to all orders in the Wilson coefficients of the EFT, similar to the so-called ``vev-improved matching'' prescription introduced in~\cite{Brehmer:2015rna}. When the 2HDM allows it, the resultant EFT is SMEFT-like in the sense that it linearly realizes electroweak symmetry, but it has a power-counting expansion determined by counting derivatives and SM fermion fields.\footnote{This combination of symmetries and power-counting is reminiscent of geoSMEFT~\cite{Helset:2020yio}, although our matching procedure incorporates higher-derivative structures that lie outside the scope of geoSMEFT (and Riemannian field-space geometry in general), and we do not organize the field dependence of Wilson coefficients geometrically.} The resummation of Higgs field dependence leads to improved convergence away from the decoupling limit. In the decoupling limit, one can of course expand the field dependence contained in these Wilson coefficients, thereby obtaining a conventional SMEFT expansion (which is understood to involve both linearly-realized electroweak symmetry and a power-counting expansion in operator dimensions). We concretely demonstrate the advantages of the SL basis by comparing the predictions for three pseudo-observables --- the Higgs coupling to gauge bosons, the Higgs self-coupling, and the Higgs coupling to fermions --- between the full theory and EFTs obtained from matching in the Higgs basis and the SL basis, finding that the SL basis generically outperforms the Higgs basis by a significant margin away from the decoupling limit.
  
The rest of this paper is organized as follows. In Section \ref{sec:bases} we begin by reviewing the general 2HDM parameterization and conditions for charge conservation. We then define the SL basis and the  transformation relating it to the Higgs basis and explore the circumstances under which each basis admits a SMEFT expansion. We carry out tree-level matching in the SL basis using functional methods in Section \ref{sec:matching}. Matching in the SL basis involves an expansion in powers of derivatives and fermions, which we carry out up to six derivatives and/or fermions, and all orders in the light Higgs doublet. Matching to all orders in the light Higgs doublet --- a feat enabled by the simplicity of the SL basis --- effectively resums zero-derivative terms in the SMEFT expansion associated with the physical masses of the heavy Higgs bosons. In Section \ref{sec:Numerics} we compare numerical predictions for key Higgs pseudo-observables between the full theory, the EFT obtained from matching in the Higgs basis, and the EFT obtained from matching in the SL basis, demonstrating the improved precision of the SL basis. We illustrate aspects of the mapping between EFTs obtained from the Higgs basis and the SL basis in Appendices \ref{app:SLToHiggs} and \ref{app:Converting}.

%%%%%%%%%%%%%%%%%%%%%%%%%%%%%%%%%%%%%%%%%%%%%%%%%%%%%%%%%%%%%%%%%%%%%%%%%%%%%%%%
\section{More Higgses, More Bases}
\label{sec:bases}
%%%%%%%%%%%%%%%%%%%%%%%%%%%%%%%%%%%%%%%%%%%%%%%%%%%%%%%%%%%%%%%%%%%%%%%%%%%%%%%%

The goal of this section is to introduce the general 2HDM and to provide a discussion of its vacuum structure. Many intricacies of the 2HDM stem from the ability to change basis by mixing the two doublets with each other.  This freedom allows us to define the straight-line (SL) basis, for which (the zero-derivative part of) the classical solution of the ``heavy'' Higgs doublet will be proportional to the ``light'' doublet. We will then provide a map between the SL basis and the Higgs basis, which will facilitate a comparison between the convergence properties of the EFTs that result when integrating out the BSM states for these two basis choices.

\subsection{Defining the 2HDM}
\label{subsec:2HDM}

The 2HDM is defined as the most general renormalizable Lagrangian built out of the Standard Model fermions and gauge bosons along with two $SU(2)_L$ doublet complex scalar fields with $U(1)_Y$ hypercharge $1/2$.  We denote them by $\Phi_a^\alpha$, together with their conjugate $\Phi^\dagger_{ a \alpha}$. There are two types of indices on the Higgs fields: a flavor index $a=1,2$ differentiates between the two doublets and the upper gauge index $\alpha$ transforms in the fundamental representation of $SU(2)_L$.

The Lagrangian comprises a set of kinetic terms (including the minimal coupling to gauge bosons through the covariant derivative $D_\mu$), the scalar potential, and Yukawa couplings,
\begin{subequations}\label{eq:2HDMLag}
\begin{align}
\lag &= \lag_2 + \lag_0 + \lag_J \,, \\[8pt]
\lag_2 &= \big( D_\mu \Phi^\dagger_{a} \big) \big(D^\mu \Phi_a \big) \,, \label{eq:2HDMkin} \\[5pt]
-\lag_0 &= Y_{a  b} \big( \Phi^\dagger_{ a} \Phi_b \big) + \frac12 Z_{a  b c  d} \big( \Phi^\dagger_{ a} \Phi_b \big) \big( \Phi^\dagger_{ c} \Phi_d \big) \,, \label{eq:2HDMpot} \\[5pt]
-\lag_J &= \mathcal{Y}^D_{i j  a} \, \overline{Q}_i d_j \Phi_a + \mathcal{Y}^{U\dagger}_{i j  a} \, \overline{u}_i (\varepsilon Q_j) \Phi_a + \mathcal{Y}^E_{i j  a} \, \overline{L}_i e_j \Phi_a + \text{h.c.} \notag\\[3pt]
&\equiv J^\dagger_{ a} \Phi_a + \text{h.c.} \,, \label{eq:2HDMyuk}
\end{align}
\end{subequations}
where we have suppressed $SU(2)_L$ gauge indices and omitted terms that are independent of the $\Phi$ fields for brevity. The $Q$, $d$, $u$, $L$, and $e$ represent the three families of Standard Model fermions. We have expressed the scalar potential in terms of the mass-dimension-2 couplings $Y_{ab}$ and dimensionless couplings $Z_{abcd}$ introduced in \cite{Branco:1999fs} (see also \cite{Davidson:2005cw, Trautner:2018ipq}), which satisfy
\begin{equation}
Y_{ab} = Y_{ba}^* \,,\qquad\quad
Z_{ab cd} = Z_{cd ab} = Z_{ba dc}^* \,.
\label{eq:YZsym}
\end{equation}
These can be related to the standard 2HDM notation,
\begin{subequations}\label{eq:mapToMsAndLambdas}
\begin{align}
\left( Y_{11}, Y_{12}, Y_{22} \right) &= \left( m_1^2, -m_{12}^2, m_2^2 \right) \,, \\[5pt]
\left( Z_{1111}, Z_{1112}, Z_{1122}, Z_{1221}, Z_{1212}, Z_{1222}, Z_{2222} \right) &= \left( \lambda_1, \lambda_6, \lambda_3, \lambda_4, \lambda_5, \lambda_7, \lambda_2 \right) \,.
\end{align}
\end{subequations}
The Yukawa matrices $\mathcal{Y}$ are \emph{a priori} arbitrary complex matrices, and together with the SM fermions they are subsumed into the $SU(2)_L$ doublet scalar currents $J_a^\alpha$, which couple to the Higgs fields.

The kinetic term $\lag_2$ is invariant under a $U(2)$ flavor symmetry,
\begin{equation}
\Phi_a \to U^\text{flavor}_{a  b} \Phi_b \,,\qquad
\text{with} \qquad U^\text{flavor} \in U(2) \,.
\label{eq:flavorTrans}
\end{equation}
Under this transformation, the couplings $Y_{a  b}$, $Z_{a  b c  d}$, and $\mathcal{Y}_{ a}$ rotate accordingly. One consequence of this freedom is that, starting from the 14 real parameters in the scalar potential (of which 4 are phases), only 11 (of which 2 phases) are physical.\footnote{Note that the central $U(1)$ subgroup of the $U(2)$, which just rephases both doublets equally, leaves the parameters invariant.}

Within the 11-dimensional physical parameter space of the scalar potential, one can identify phenomenologically viable subspaces. Requiring \emph{explicit CP conservation} amounts to turning off the 2 physical phases, which is equivalent to demanding that there exists a basis, accessed by flavor rotations, where all $Y_{ab}$ and $Z_{abcd}$ parameters are real \cite{Gunion:2005ja}. If we further require \emph{explicit custodial symmetry conservation} in the scalar potential, then this is equivalent to further requiring $Z_{1221} = Z_{1212}$ in the basis with real valued $Y_{ab}$ and $Z_{abcd}$ \cite{Haber:2010bw} (see also \cite{Pomarol:1993mu}). In spite of their explicit conservation, CP and custodial symmetry may yet be spontaneously broken by the vacuum configuration of the two Higgses.

Electric charge can be spontaneously broken by the vacuum configuration of the 2HDM with or without explicit CP conservation. It is understood that a vacuum configuration conserves charge if and only if a unitary gauge rotation can be found to simultaneously set the upper components of both Higgs vevs to zero~\cite{Ferreira:2004yd, Barroso:2005sm},
\begin{equation}
\Phi_1 \bigr|_\text{vev} = \frac{1}{\sqrt{2}} \mqty( 0 \\ v_1 ) \,,\qquad
\Phi_2 \bigr|_\text{vev} = \frac{1}{\sqrt{2}} \mqty( 0 \\ v_2 ) \,.
\end{equation}
We introduce a complex number (assuming w.l.o.g.\ that $v_1$ is real)
\begin{equation}
k \equiv \frac{v_2}{v_1} \in \mathbb{C} \,,
\end{equation}
where $\abs{k} = \tan \beta$.
This allows us to recast the above criterion in a general gauge basis as the requirement that the two Higgs vevs are multiples of each other (\ie\ aligned in the gauge space),
\begin{equation}
\Phi_2^\alpha \bigr|_\text{vev} = k\s \Phi_1^\alpha \bigr|_\text{vev} \,.
\end{equation}
On phenomenological grounds, we work with the 2HDM parameter space for which this criterion is satisfied. Note that if this criterion is satisfied in one flavor basis, it is satisfied in any flavor basis, but the ratio $k$ is different in different bases.

\subsection{The Straight-line Basis}
\label{subsec:SLBasis}

We assume that we are working in a region of parameter space where the BSM Higgs states are sufficiently heavy for it to be useful to integrate them out.  There then exists a direction in flavor space such that the second Higgs doublet $\Phi_2$ is ``heavy,'' meaning that its components are sufficiently well aligned with the larger eigendirections of the mass matrix at the global minimum.  Our goal is then to integrate out $\Phi_2$ in order to obtain an EFT describing the low energy behavior of the ``light'' doublet $\Phi_1$. Here, we employ the functional approach for matching onto the EFT by integrating out the heavy states in the path integral in the semiclassical approximation (see~\cite{Cohen:2022tir, Dawson:2022ewj} for recent reviews of functional matching and implementation).  At tree level this amounts to finding a classical solution to the equations of motion for the heavy doublet, $\Phi_{2,\textbf{c}}[\Phi_1]$, and substituting it back into the 2HDM action to yield the tree-level EFT. This generates the EFT operators and their Wilson coefficients together and facilitates working to all orders in the field $\Phi_1$.

Working order-by-order in powers of derivatives, we require that the zero-derivative part of the classical solution,
\begin{equation}
\Phi_{2,\textbf{c}} [\Phi_1] = \Phi_{2,\textbf{c}}^{(0)} (\Phi_1) + O\big(\partial^2\big) \,,
\end{equation}
solves the zero-derivative part of $\Phi_2$'s equation of motion, namely,
\begin{equation}
-\frac{\partial \lag_0}{\partial \Phi_2^\dagger}\bigg|_{\Phi_2 = \Phi_{2,\textbf{c}}^{(0)}(\Phi_1)}
= Y_{2 b} \Phi_b \bigr|_{\Phi_2 = \Phi_{2,\textbf{c}}^{(0)}(\Phi_1)}
+ Z_{2 b c d}\s \Phi_b \big( \Phi^\dagger_c \Phi_d \big) \bigr|_{\Phi_2 = \Phi_{2,\textbf{c}}^{(0)}(\Phi_1)} = 0 \,. 
\label{eq:zeroDerEOM}
\end{equation}
This is a cubic equation in $\Phi_2$; in a generic 2HDM basis it yields an EFT submanifold curve $\Phi_{2,\textbf{c}}^{(0)}(\Phi_1)$ that is a complicated function. Now we will show that one can find a special 2HDM basis in which the solution curve $\Phi_{2,\textbf{c}}^{(0)}(\Phi_1)$ is simply a straight line as long as the 2HDM has a global minimum that preserves electric charge. We refer to this basis as the SL basis.

We begin by noting that \cref{eq:zeroDerEOM} must be satisfied at the point corresponding to the global minimum because by definition this is a point that minimizes the potential,
\begin{equation}
Y_{2 b} \Phi_b \bigr|_\text{vev}
+ Z_{2 b c d}\s \Phi_b \big( \Phi^\dagger_c \Phi_d \big) \bigr|_\text{vev} = 0 \,. 
\label{eq:EOMvev}
\end{equation}
Let us focus on the first term. The key observation is that it is the lower component of the ``vector''
\begin{equation}
Y_{ab} \Phi_b \bigr|_\text{vev} \,,
\end{equation}
which transforms in the fundamental representation of the flavor rotation group in \cref{eq:flavorTrans}. Therefore, one can always find a flavor basis such that its lower component vanishes,
\begin{equation}
Y_{2b} \Phi_b \bigr|_\text{vev} = 0 \qquad\quad \text{(SL basis condition)} \,.
\label{eq:SLBasisCondition}
\end{equation}
This defines our SL basis, in which the two terms in \cref{eq:EOMvev} both vanish independently,
\begin{equation}
Y_{2 b} \Phi_b \bigr|_\text{vev} = Z_{2 b c d}\s \Phi_b \big( \Phi^\dagger_c \Phi_d \big) \bigr|_\text{vev} = 0 \,.
\label{eqn:bothvanish} 
\end{equation}
Note that if a homogeneous function of $\Phi_a$ vanishes at a certain charge-conserving point (where their values are multiples of each other), then it vanishes on the whole (charge-conserving) straight line that connects that point with the origin. Since the two terms in \cref{eq:zeroDerEOM} are both homogeneous functions of $\Phi_a$, \cref{eqn:bothvanish} implies that they both also vanish on the straight line,
\begin{equation}
Y_{2 b} \Phi_b \bigr|_{\Phi_2 = k\s \Phi_1} = Z_{2 b c d}\s \Phi_b \big( \Phi^\dagger_c \Phi_d \big) \bigr|_{\Phi_2 = k\s \Phi_1} = 0 \,. 
\end{equation}
Therefore, in the SL basis the EOM \cref{eq:zeroDerEOM} has the straight line solution
\begin{equation}
\Phi_{2,\textbf{c}}^{(0)}(\Phi_1) = k\s \Phi_1 \,,
\qquad \text{with} \qquad
k \equiv \frac{v_2}{v_1} \in \mathbb{C}
\qquad \text{in the SL basis.}
\label{eq:SLbasis}
\end{equation}

Although in the SL basis \cref{eq:SLbasis} is always a solution to the EOM in \cref{eq:zeroDerEOM}, this straight-line EFT submanifold can only correspond to a well-behaved SMEFT when $Y_{22}>0$ in the SL basis; see \cref{sec:SMEFTvsHEFT}.

\subsection{Mapping Between the Straight-line and Higgs Bases}
\label{subsec:mappingSLHiggs}

Let us write the doublets in the SL basis as $\Phi_a$ ($a=1,2$), and the doublets in the Higgs basis as $\Phi_\hg{a}$ ($\hg{a}=\hg{1},\hg{2}$), adopting a convention of \textit{dotting} Higgs-basis indices. We seek the unitary matrix $U_{\hg{a}b}$ that relates the two,
\begin{equation}
\Phi_\hg{a} = U_{\hg{a}b}\s \Phi_b \,.
\end{equation}
The vevs in the two bases are similarly related,
\begin{equation}
v_\hg{a} = U_{\hg{a}b}\s v_b \,.
\end{equation}
As the vevs in the respective bases are defined as
\begin{equation}
v_\hg{a} = \mqty( v \\ 0 ) \,,\qquad
v_a = \mqty( v_1 \\ v_2 ) = \frac{v}{\sqrt{1+\abs{k}^2}} \mqty( 1 \\ k ) \,,
\end{equation}
where $v^2 = v_1^2 + \abs{v_2}^2$, it follows that
\begin{equation}
U_{\hg{a}b} = \frac{1}{\sqrt{1+\abs{k}^2}} \mqty( 1 & k^* \\ -k & 1 ) \,.
\label{eq:Udef}
\end{equation}

Rearranging the definition of the SL basis in \cref{eq:SLBasisCondition} allows us to define $k$ in terms of quadratic pieces of the SL basis potential,
\begin{equation}
k = \frac{v_2}{v_1} = -\frac{Y_{21}}{Y_{22}} \,.
\label{eq:kAsSLParams}
\end{equation}
As $U$ relates the quadratic parameters in the SL and Higgs bases via
\begin{equation}
Y_{\hg{a} \hg{b}} = U_{\hg{a}c}\, Y_{c d}\, U^\dagger_{d \hg{b}} \,,
\end{equation}
$k$ can also be written in terms of Higgs basis quantities,
\begin{equation}
-k = \frac{Y_{\hg{2}\hg{1}}}{Y_{\hg{1}\hg{1}}} = \frac{Z_{\hg{2}\hg{1}\hg{1}\hg{1}}}{Z_{\hg{1}\hg{1}\hg{1}\hg{1}}} \,.
\label{eq:kAsHParams}
\end{equation}
The last equality comes from the vev conditions in the Higgs basis, which relate
\begin{equation}
-v^2 =  \frac{2 Y_{\hg{1}\hg{1}}}{Z_{\hg{1}\hg{1}\hg{1}\hg{1}}} = \frac{2 Y_{\hg{2}\hg{1}}}{Z_{\hg{2}\hg{1}\hg{1}\hg{1}}} \,.
\label{eq:HBVevConditions}
\end{equation}
The map between other SL and Higgs basis quantities that appear in the EFT matching is provided in \cref{app:SLToHiggs}.

We note that both the SL and the Higgs basis are actually a $U(1)$ family of bases. This corresponds to the freedom to rephase the second Higgs doublet, without affecting the respective bases' vev conditions of \cref{eq:SLBasisCondition} and $v_\hg{2}=0$. The above procedure details a one-to-one map between equivalent SL and Higgs bases. This means that real scalar potential parameters unaffected by this rephasing --- in the SL basis as in the Higgs basis --- are physical.\footnote{We thank H.\ Haber for pointing this out.}

\subsection{Matching Onto SMEFT or HEFT}
\label{sec:SMEFTvsHEFT}

As we emphasized in the introduction, SMEFT is the EFT extension of the Standard Model that is expressed about the origin in field space where $\left\lvert\Phi_1 \right\rvert = 0$ such that electroweak symmetry can be linearly realized.  For SMEFT to be well defined, the EFT must be built from analytic functions of $\Phi_1$, which admit a convergent expansion of local operators at this point.  If it is not, then the UV theory must be matched onto HEFT \cite{Cohen:2020xca}.  This invites the question: is it possible to determine which regions of the 2HDM parameter space can be matched onto SMEFT?

\begin{figure}
\centering
\includegraphics[width=0.55\textwidth]{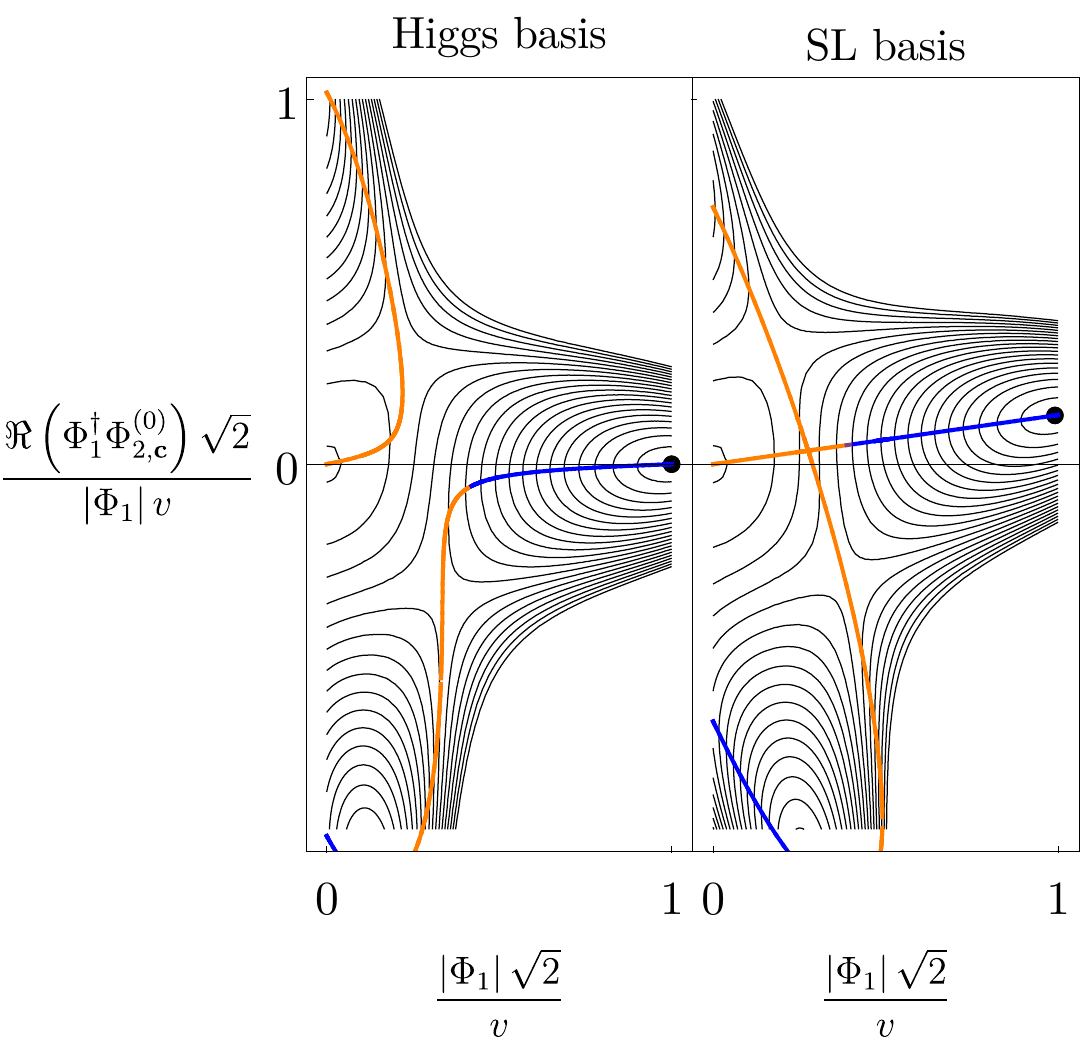}\\[10pt]
\includegraphics[width=0.55\textwidth]{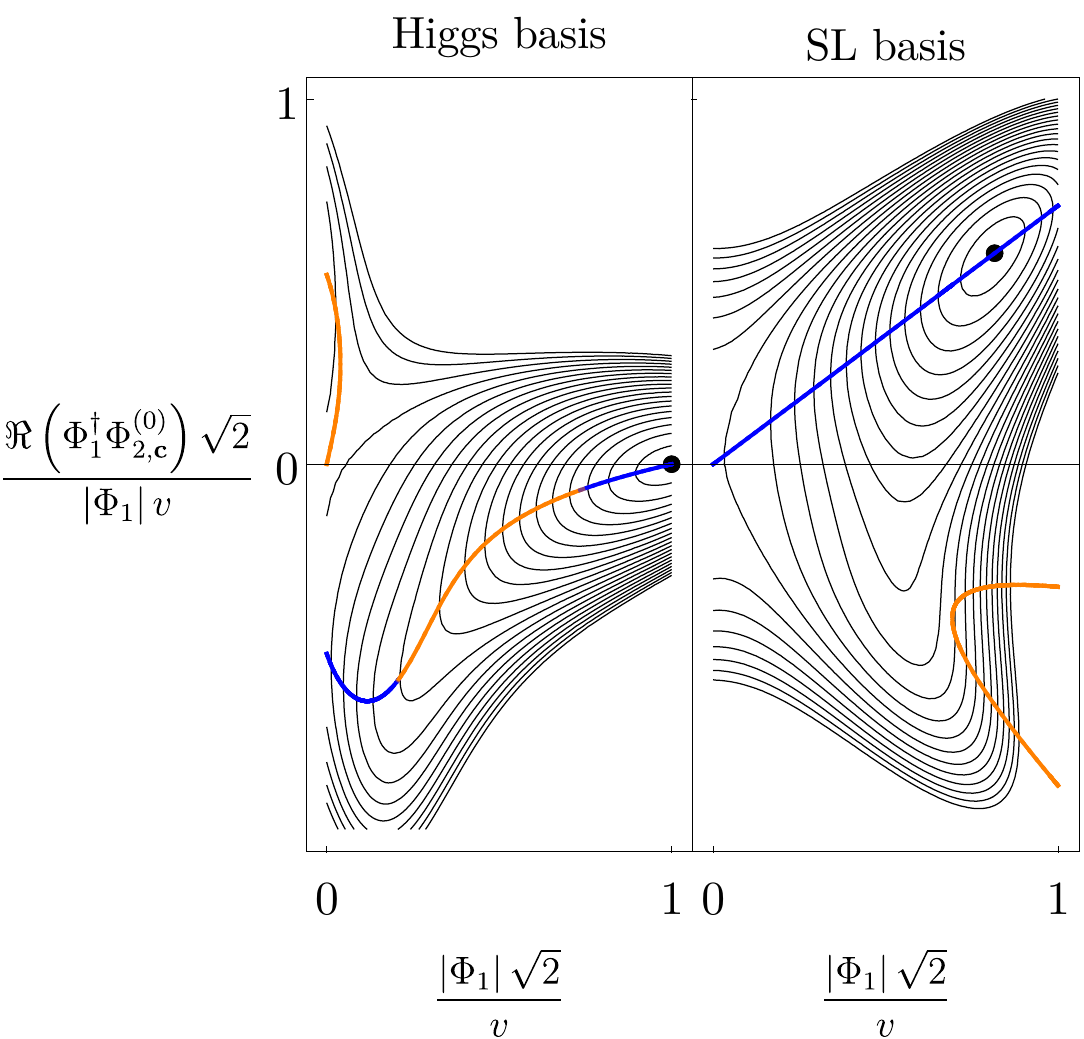}
\caption{The Higgs and SL basis behavior for two example custodially symmetric 2HDM models. Black contours show the potential, and a black dot shows the global minimum, which has coordinates $(\cos\beta,\sin\beta)$ on these axes. The zero-derivative solutions of $\Phi_2$'s EOM, $\Phi_{2,\textbf{c}}^{(0)}$, are shown in blue if the $\Phi_2$ mass matrix is positive definite and in orange otherwise. Top: an example where $Y_{22}<0$ in both bases, and neither matches onto SMEFT. Bottom: $Y_{22}>0$ in the SL basis, whereas $Y_{\hg{2}\hg{2}}<0$ in the Higgs basis. \label{fig:SMEFTHEFTExamples}}
\end{figure}

\cref{fig:SMEFTHEFTExamples} visualizes the charge conserving solutions $\Phi_{2,\textbf{c}}^{(0)} (\Phi_1)$ of $\Phi_2$'s zero-derivative EOM, \cref{eq:zeroDerEOM}, by plotting
\begin{equation}
\frac{\Re \left( \Phi_1^\dagger \Phi_{2,\textbf{c}}^{(0)} \right) \sqrt{2}}{\left| \Phi_1 \right| v} \quad\text{versus}\quad \frac{\left| \Phi_1 \right| \sqrt{2}}{v} \,.
\end{equation}
In these coordinates, the global minimum lies at $(\cos \beta,\sin\beta)$, and in the SL basis, one of the solutions is a straight line of gradient $\Re k$. \cref{fig:SMEFTHEFTExamples} shows two different custodially symmetric UV parameter points in both their respective Higgs and SL bases; custodial symmetry guarantees that $\Im \left( \Phi_1^\dagger \Phi_{2,\textbf{c}}^{(0)} \right) = 0$ and $\Im k = 0$. Black contours show the 2HDM potential in the space of $\Phi_1$ and $\Phi_{2,\textbf{c}}^{(0)}$. The global minimum is shown by a black dot. The potential contours and global minimum are rotated between the Higgs and SL bases.

The multiple solutions for $\Phi_{2,\textbf{c}}^{(0)}$ are the paths that extremize the potential in the vertical direction. The solutions shown in blue are stable --- the mass matrix of the $\Phi_2$ modes (\cref{eq:MassMat}) about blue solutions is positive definite; those shown in orange are not. Notably, the solutions $\Phi_{2,\textbf{c}}^{(0)}$ in the Higgs and SL basis EFTs are not simple rotations of each other. Even when starting from the same UV parameter point, the resulting Higgs and SL basis EFTs are generally different (truncated to zero derivative order) and are not both guaranteed to admit a SMEFT expansion.

Following the treatment of \cite{Cohen:2020xca}, consider the behavior of the EFTs in the $\left\lvert\Phi_1 \right\rvert \to  0$ limit. We will first argue that $Y_{22} <0$ is a sufficient criterion for a given basis' EFT not to match on to SMEFT. In the SL basis, $Y_{22} <0$ leads to tachyonic modes in $\Phi_2$'s mass matrix about the solution $\Phi_{2,\textbf{c}}^{(0)}$, \cref{eq:MassMat}, in the $\abs{\Phi_1} \to 0$ limit. This EFT does not have a region of small $p^2$ where the effects of $\Phi_2$ are purely virtual. The sickness is most apparent when matching at loop level: when $\left\lvert\Phi_1 \right\rvert \to  0$ the Lagrangian would have an anti-Hermitian component corresponding to a rate for tunneling out of the false vacuum $\Phi_{2,\textbf{c}}^{(0)}$.

In the Higgs basis, $Y_{\hg{2}\hg{2}} < 0$ generally leads to $\Phi_{2,\textbf{c}}^{(0)}$ approaching a non-zero constant as $\left\lvert\Phi_1 \right\rvert \to  0$. This does not yield a SMEFT, as can be verified by substituting $\Phi_{2,\textbf{c}}^{(0)}$ back into the kinetic term \cref{eq:2HDMkin}. As $\left\lvert\Phi_1 \right\rvert \to  0$, the $W$ mass remains non-zero, which cannot be reproduced using SMEFT operators.

Of course, whether $Y_{22} < 0$ can be a \emph{basis dependent} statement. If both eigenvalues of the matrix $Y_{ab}$ are negative, then $Y_{22} < 0$ is guaranteed in both the Higgs and SL bases, as is the case for the potential in the top half of \cref{fig:SMEFTHEFTExamples}. However, if only one eigenvalue of $Y_{ab}$ is negative, the sign of $Y_{22}$ may vary among bases. In this case, \cref{eqn:Y22inHiggs} guarantees that in the SL basis,
\begin{equation}
Y_{22} = \frac{1}{1+|k|^2} \frac{\det Y}{Y_{\hg{1}\hg{1}}} > 0 \,,
\end{equation}
and the SL basis EFT formally admits a SMEFT expansion. (It is nonetheless possible that the SMEFT expansion does not converge at the global minimum and therefore has no predictive power for low energy observables we might measure. This will happen if the inverse of the mass scale, defined in \cref{eq:SLMassScale}, when viewed as a function of $v$, has a radius of convergence about $v=0$ that is smaller than the true value of $v= 246 \, \mathrm{GeV}$ \cite{Cohen:2020xca}.)

Even if $Y_{22} > 0$ in the SL basis, it is possible that, simultaneously, $Y_{\hg{2}\hg{2}} < 0$ in the Higgs basis, as shown in the bottom example of \cref{fig:SMEFTHEFTExamples}. Thus, working in the SL basis improves the chances of matching onto SMEFT as $Y_{22} > 0$ whenever possible. As we will see in \cref{sec:Numerics}, working in the SL basis also improves the convergence of the resulting EFT expansion.

%%%%%%%%%%%%%%%%%%%%%%%%%%%%%%%%%%%%%%%%%%%%%%%%%%%%%%%%%%%%%%%%%%%%%%%%%%%%%%%%
\section{Matching in the SL Basis}
\label{sec:matching}
%%%%%%%%%%%%%%%%%%%%%%%%%%%%%%%%%%%%%%%%%%%%%%%%%%%%%%%%%%%%%%%%%%%%%%%%%%%%%%%%

We will now use the classical solution to the equation of motion for the second Higgs doublet in the SL basis to integrate out $\Phi_2$ at tree level.  We will include terms in the EFT up to six derivative and/or fermion order and to all orders in the light field $\Phi_1$.

\subsection{Organizing the EFT Expansion}
\label{subsec:organizing}

Since we need to derive terms involving as many as six derivatives and/or fermions in the EFT, we begin by setting up the expansion of the UV action on the classical equations of motion for the heavy doublet.
We write the UV action derived using the Lagrangian in Eqs.~(\ref{eq:2HDMLag}) as
\begin{equation}
  S_\text{UV}[\Phi_2] = S_0[\Phi_2] + \epsilon \big(  S_2[\Phi_2] + S_J[\Phi_2] \big) = S_0[\Phi_2] + \epsilon \s S_\epsilon[\Phi_2] \,,
  \label{eq:SUV}
\end{equation}
where $S_\epsilon[\Phi_2]$ is implicitly defined here, $S_0[\Phi_2]$ contains the zero-derivative scalar terms, $S_2[\Phi_2]$ contains the two-derivative scalar terms, $S_J[\Phi_2]$ contains the Yukawa interactions, and $\epsilon$ is an order parameter which we use to track the sum of the number of fermions and derivatives,
\begin{equation}
  2\epsilon = \# \text{ of derivatives} + \# \text{ of fermions} \,.
\end{equation}
Ultimately, we will set $\epsilon=1$.  Note that we are only writing the explicit functional dependence on $\Phi_2$ here for brevity, but of course $S_\text{UV}$ also depends on the light Standard Model fields.  

We will denote the Higgs doublet we are integrating out at tree level as
\begin{equation}
\mathcal{H}_x = \mqty( \Phi_2^\alpha(x) \\[3pt] \Phi^\dagger_{2\alpha}(x) ) \,,
\end{equation}
where the $x$ label simultaneously stands for 1) the spacetime coordinate, 2) the $SU(2)_L$ index, and 3) the Higgs doublet versus its conjugate, as we need to vary with respect to all of them.
We want to find $\mathcal{H}_{\mathbf{c},x}$, the classical solution to the equation of motion for $\mathcal{H}_x$, order-by-order in $\epsilon$,
\begin{equation}
\mathcal{H}_{\mathbf{c},x} = \sum_{n=0}^\infty  \epsilon^{n}\, \HH{x}{n} \,.
\label{eq:PhiExpansion}
\end{equation}
This allows us to derive the EFT action as a semiclassical expansion,
\begin{equation}
S_\text{EFT}^\text{tree} = S_\text{UV}[\mathcal{H}_{\mathbf{c},x}] \,.
\end{equation}
Substituting the expansion defined in \cref{eq:PhiExpansion} into \cref{eq:SUV}, we find
\begin{align}
S_\text{EFT}^\text{tree} &= \epsilon^0\s \overline{S_0}
\notag\\[5pt]
&\hspace{10pt}
+ \epsilon^1 \Big[\s \overline{S_\epsilon} + (\overline{\delta S_0})_x\, \HH{x}{1} \s\Big]
\notag\\[5pt]
&\hspace{10pt}
+ \epsilon^2 \bigg[\s \frac12\, (\overline{\delta^2 S_0})_{xy}\, \HH{x}{1}\, \HH{y}{1} + (\overline{\delta S_\epsilon})_x\, \HH{x}{1} +  (\overline{\delta S_0})_x\, \HH{x}{2} \s\bigg]
\notag\\[5pt]
&\hspace{10pt}
+ \epsilon^3 \bigg[\s \frac12\, (\overline{\delta^2 S_\epsilon})_{xy}\, \HH{x}{1}\, \HH{y}{1} + \frac16\, (\overline{\delta^3 S_0})_{xyz}\, \HH{x}{1}\, \HH{y}{1}\, \HH{z}{1}
\notag\\[0pt]
&\hspace{40pt}
+ (\overline{\delta^2 S_0})_{xy}\, \HH{x}{1}\, \HH{y}{2} + (\overline{\delta S_\epsilon})_x\, \HH{x}{2} +  (\overline{\delta S_0})_x\, \HH{x}{3} \s\bigg]
\notag\\[5pt]
&\hspace{10pt}
+ O\big(\epsilon^4\big) \,.
\label{eq:UnsimpEFTS}
\end{align}
We use a bar to denote quantities evaluated on the zeroth-order classical solution $\mathcal{H}_{\mathbf{c},x}^{(0)}$, and we have defined the shorthand
\begin{equation}
(\delta S)_x\equiv\frac{\delta S}{\delta \mathcal{H}_x} \,,\qquad
(\delta^2S)_{xy}\equiv\frac{\delta^2 S}{\delta \mathcal{H}_x \delta \mathcal{H}_y} \,,\qquad
(\delta^3S)_{xyz}\equiv\frac{\delta^3 S}{\delta \mathcal{H}_x\delta \mathcal{H}_y\delta \mathcal{H}_z} \,.
\end{equation}
Note that a repeated index implies an integral over the associated spacetime coordinate as well as a sum over the components of the Higgs doublet and their conjugates.

To find $\mathcal{H}_{\mathbf{c},x}$, we expand the equation of motion in powers of $\epsilon\,$,
\begin{align}
0 &= \frac{\delta S_\text{UV}}{\delta \mathcal{H}_x} \bigg\vert_{\mathcal{H}_x=\mathcal{H}_{\mathbf{c},x}}
\notag\\[8pt]
&= \epsilon^0\, (\overline{\delta S_0})_x + \epsilon^1 \Big[\s (\overline{\delta S_\epsilon})_x + (\overline{\delta^2 S_0})_{xy}\, \HH{y}{1} \s\Big]
\notag\\[5pt]
&\hspace{10pt} + \epsilon^2 \bigg[\s (\overline{\delta^2 S_\epsilon})_{xy}\, \HH{y}{1} + (\overline{\delta^2 S_0})_{xy}\, \HH{y}{2} + \frac12\, (\overline{\delta^3 S_0})_{xyz}\, \HH{y}{1}\, \HH{z}{1} \s\bigg] + O\big(\epsilon^3\big) \,.
\end{align}
Each order in $\epsilon$ must independently be zero. This gives
\begin{subequations}\label{eq:allorder}
\begin{align}
(\overline{\delta S_0})_x &= 0\,, \label{eq:zerothorder} \\[8pt]
(\overline{\delta S_\epsilon})_x + (\overline{\delta^2 S_0})_{xy}\, \HH{y}{1} &=0 \,, \label{eq:firstorder} \\[3pt]
(\overline{\delta^2 S_\epsilon})_{xy}\, \HH{y}{1} + (\overline{\delta^2 S_0})_{xy}\, \HH{y}{2} + \frac12 (\overline{\delta^3 S_0})_{xyz}\, \HH{y}{1}\, \HH{z}{1} &= 0 \,, \label{eq:secondorder}
\end{align}
\end{subequations}
which can be solved to give $\mathcal{H}_{\mathbf{c},x}$ order-by-order in $\epsilon$. Note that Eqs.~(\ref{eq:allorder}) imply an immediate simplification of \cref{eq:UnsimpEFTS},
\begin{align}
S_\text{EFT}^\text{tree} &= \epsilon^0\, \overline{S_0} + \epsilon^1\, \overline{S_\epsilon}
+ \epsilon^2 \bigg[ -\frac12\, (\overline{\delta^2S_0})_{xy}\, \HH{x}{1}\, \HH{y}{1} \s\bigg]
\notag\\[5pt]
&\hspace{10pt}
+ \epsilon^3 \bigg[\s \frac12\, (\overline{\delta^2S_\epsilon})_{xy}\, \HH{x}{1}\, \HH{y}{1} + \frac16\, (\overline{\delta^3S_0})_{xyz}\, \HH{x}{1}\, \HH{y}{1}\, \HH{z}{1} \s\bigg] \,. 
\label{eq:SimpEFTS}
\end{align}
We thus only need to compute $\mathcal{H}_{\mathbf{c},x}^{(1)}$, which amounts to solving \cref{eq:firstorder}. This requires  inverting the mass matrix $(\overline{\delta^2S_0})_{xy}$, as described in the next section.

\subsection{Inverting the Mass Matrix}

The general expansion derived in the preceding subsection is valid in a general flavor basis. As mentioned above, deriving the EFT to the desired order requires solving \cref{eq:firstorder}.  We therefore must invert the mass matrix.  To do so, we now specialize to the SL basis as defined in \cref{eq:SLbasis}, for which
\begin{align}
\! (\overline{\delta^2 S_0})_{xy} =\! - \delta^{(4)}(x-y)\! \begin{pmatrix}
  Z_1 \Phi^\dagger_{1 \alpha} \Phi^\dagger_{1 \beta} & (Y_{22}+Z_2 |\Phi_1|^2) \delta_\alpha^\beta + Z_3 \Phi^\dagger_{1 \alpha} \Phi^\beta_1 \\[5pt]
   (Y_{22}+Z_2 |\Phi_1|^2) \delta^\alpha_\beta + Z_3 \Phi^\dagger_{1 \beta} \Phi^\alpha_1 & Z_1^* \Phi_1^\alpha \Phi_1^\beta
\end{pmatrix} , \nonumber\\
\label{eq:MassMat}
\end{align}
where
\begin{subequations}
\begin{align}
Z_1 &= Z_{1212} + 2 k^* Z_{1222} + (k^*)^2 Z_{2222} =
   \begin{pmatrix}
    1 & k^*
  \end{pmatrix}
\begin{pmatrix}
  Z_{1212} & Z_{1222} \\ Z_{1222} & Z_{2222}
\end{pmatrix}
  \begin{pmatrix}
    1 \\ k^*
  \end{pmatrix} \,, \\[5pt]
Z_2 &= Z_{1122} + 2 \Re\bigl[ k Z_{1222} \bigr] + |k|^2 Z_{2222} =
    \begin{pmatrix}
      1 & k^*
    \end{pmatrix}
  \begin{pmatrix}
    Z_{1122} & Z_{1222} \\ Z_{2122} & Z_{2222}
  \end{pmatrix}
    \begin{pmatrix}
      1 \\ k
    \end{pmatrix} \,, \\[5pt]
Z_3 &= Z_{1221} + 2 \Re\bigl[ k Z_{1222} \bigr] + |k|^2 Z_{2222} = \begin{pmatrix}
      1 & k^*
    \end{pmatrix}
  \begin{pmatrix}
    Z_{1221} & Z_{1222} \\ Z_{2122} & Z_{2222}
  \end{pmatrix}
    \begin{pmatrix}
      1 \\ k
    \end{pmatrix} \,.
\end{align}
\end{subequations}
Note that $Z_2$ and $Z_3$ are real valued. Consistency with the $SU(2)_L$ structure implies an ansatz for the inverse,
\begin{equation}
(\overline{\delta^2 S_0})^{-1}_{yz} = - \delta^{(4)}(y-z) \begin{pmatrix}
A\, \Phi_1^\beta \Phi_1^\gamma & B\, \delta^\beta_\gamma + C\, \Phi^\dagger_{1 \gamma} \Phi^\beta_1 \\[5pt]
B\, \delta_\beta^\gamma + C\, \Phi^\dagger_{1 \beta} \Phi^\gamma_1 & A^*\, \Phi^\dagger_{1\beta} \Phi^\dagger_{1\gamma}
\end{pmatrix} \,,
\label{eq:invMassMat}
\end{equation}
with $B=B^*$ and $C=C^*$. The solution is given by
\begin{subequations}\label{eq:straightLineInverseMassMatCoeffs}
\begin{align}
A &= -\frac{Z_1^*}{ \big[ Y_{22} + (Z_2+Z_3) |\Phi_1|^2 \big]^2 - |Z_1|^2 |\Phi_1|^4 } \,, \\[8pt]
B &=  \frac{1}{ Y_{22} + Z_2 |\Phi_1|^2 } \,, \\[8pt]
C &= -\frac{1}{ Y_{22} + Z_2 |\Phi_1|^2 } \frac{ Z_3 \big[ Y_{22} + (Z_2+Z_3) |\Phi_1|^2 \big] - |Z_1|^2 |\Phi_1|^2 }{ \big[ Y_{22} + (Z_2+Z_3) |\Phi_1|^2 \big]^2 - |Z_1|^2 |\Phi_1|^4 } \,,
\end{align}
\end{subequations}
as can be checked by explicit matrix multiplication.
With the result in \cref{eq:invMassMat}, we obtain an $O(\epsilon)$ solution to the EOM,
\begin{equation}
\Phi_{2,\textbf{c}}^{(1)} = - \left[ A \left( \Phi_1^\dagger\R \right)^* +C \left( \Phi_1^\dagger\R \right) \right] \Phi_1 - B\R \,,\quad\text{with}\quad
\R \equiv k D^2 \Phi_1 + J_2 \,.
\label{eqn:Phi2NLO}
\end{equation}

These results simplify in the custodial limit, for which, without loss of generality, all potential parameters and therefore $k$ are real, and $Z_{1221}=Z_{1212}$~\cite{Haber:2010bw} (implying $Z_1=Z_3$). The coefficients of the inverse mass matrix defined in Eqs.~(\ref{eq:straightLineInverseMassMatCoeffs}) therefore simplify to
\begin{subequations}
\begin{align}
A= C &= -\frac{Z_1}{(Y_{22}+Z_2 |\Phi_1|^2)\left[ Y_{22}+(Z_2+2 Z_1) |\Phi_1|^2 \right]} \,, \\[5pt]
B &= \frac{1}{Y_{22}+Z_2 |\Phi_1|^2} \,,
\end{align}
\end{subequations}
when the UV 2HDM respects custodial symmetry.

\subsection{The EFT Result}

We now have everything we need to determine an EFT action for the light doublet $\Phi_1$. Combining \cref{eq:SimpEFTS} with \cref{eq:SLbasis,eqn:Phi2NLO}, we have
\begin{align}
\lag_\text{EFT} &= - \left( 1+|k|^2 \right) m_\text{eff}^2 \left| \Phi_1 \right|^2 - \frac12 \left( 1+|k|^2 \right)^2 \lambda_\text{eff} \left| \Phi_1 \right|^4
\notag\\[5pt]
&\hspace{15pt}
+ \left( 1+|k|^2 \right) \left| D_\mu \Phi_1 \right|^2 - \left[ \left( J_1^\dagger + k J_2^\dagger \right) \Phi_1 + \text{h.c.} \right]
\notag\\[5pt]
&\hspace{15pt}
+ B \left| \R \right|^2 + C \big| \Phi_1^\dagger \R \big|^2 + \frac12 \left[ A^* \left( \Phi_1^\dagger \R \right)^2 + \text{h.c.} \right] + \left| D_\mu \Phi_{2,\textbf{c}}^{(1)} \right|^2 \notag\\[5pt]
&\hspace{15pt} - \left[ Z_4 \left( \Phi_1^\dagger \Phi_{2,\textbf{c}}^{(1)} \right) \left| \Phi_{2,\textbf{c}}^{(1)} \right|^2 + \text{h.c.} \right] \,,
\label{eqn:LagEFTPhi1}
\end{align}
where $A, B, C$ are given in Eqs.~(\ref{eq:straightLineInverseMassMatCoeffs}); $\Phi_{2,\textbf{c}}^{(1)}$ and $\R$ are given in \cref{eqn:Phi2NLO}. We have also introduced the notation $m_\text{eff}^2$ and $\lambda_\text{eff}$,
\begin{subequations}
\begin{align}
\left( 1 + |k|^2 \right) m_\text{eff}^2 &= Y_{ab} \begin{pmatrix} 1 \\ k^* \end{pmatrix}_a \begin{pmatrix} 1 \\ k \end{pmatrix}_b \,, \\[8pt]
\left( 1 + |k|^2 \right)^2 \lambda_\text{eff} &= Z_{abcd} \begin{pmatrix} 1 \\ k^* \end{pmatrix}_a \begin{pmatrix} 1 \\ k \end{pmatrix}_b \begin{pmatrix} 1 \\ k^* \end{pmatrix}_c \begin{pmatrix} 1 \\ k \end{pmatrix}_d \,,
\end{align}
\end{subequations}
as well as $Z_4$, which populates the elements of $(\overline{\delta^3S_0})_{xyz}$,
\begin{equation}
  Z_4 =   Z_{1222} + k^* Z_{2222} =
  \begin{pmatrix}
   1 & k^*
 \end{pmatrix}
\begin{pmatrix}
 Z_{1222} \\ Z_{2222}
\end{pmatrix} \, .
\end{equation}

\subsection{EFT Predictions for Benchmark Pseudo-observables}
\label{subsec:predictions}

We will use the matching result \cref{eqn:LagEFTPhi1} to compute three pseudo-observables: the shift in the $h\s W^+ W^-$ coupling relative to the Standard Model $\kappa_V$, the shift in the Higgs self-coupling $h^3$ relative to the Standard Model $\kappa_\lambda$, and the shift in the $h\bar{f}f$ coupling relative to the Standard Model $\kappa_f$. We will compute all of these to leading non-trivial order.

We can drop the last line of \cref{eqn:LagEFTPhi1} --- which originates from the second $O\big(\epsilon^3\big)$ term in \cref{eq:SimpEFTS} --- because it does not contribute to our pseudo-observables at the truncation order imposed in this section. Note also that the kinetic term for $\Phi_1$ is not canonically normalized. Rewriting with the normalized field, 
\begin{equation}
H \equiv \left( 1 + |k|^2 \right)^{1/2} \Phi_1 \,,
\label{eqn:normalizing}
\end{equation}
we get
\begin{align}
\lag_\text{EFT} &\supset \left| D_\mu H \right|^2 - m_\text{eff}^2 \left| H \right|^2 - \frac12 \lambda_\text{eff} \left| H \right|^4 - \left(1+|k|^2\right)^{-1} \left[ \big( \hat{J}_1^\dagger + k\s \hat{J}_2^\dagger \big) H + \text{h.c.} \right] 
\notag\\[5pt]
&\hspace{10pt}
+ \hat{B} \big| \Rhat \big|^2 + \hat{C} \big| H^\dagger \Rhat \big|^2 + \frac12 \left[ \hat{A}^* \big( H^\dagger \Rhat \big)^2 + \text{h.c.} \right] + \left(1+|k|^2\right) \left| D_\mu \hat{\Phi}_{2,\textbf{c}}^{(1)} \right|^2 \,,
\label{eq:relevantTermsInEFT}
\end{align}
with a variety of rescaled quantities,
\begin{subequations}
\begin{align}
\hat{Y}_{22} &\equiv \left(1+|k|^2\right) Y_{22} \,, \\[10pt]
\hat{J}_i &\equiv \left(1+|k|^2\right)^{1/2} J_i \,, \\[10pt]
\hat{A} &\equiv \left(1+|k|^2\right)^{-2} A = - \frac{Z_1^*}{ \big[ \hat{Y}_{22} + (Z_2+Z_3) |H|^2 \big]^2 - |Z_1|^2 |H|^4 } \,, \\[5pt]
\hat{B} &\equiv \left(1+|k|^2\right)^{-1} B = \frac{1}{\hat{Y}_{22} + Z_2 |H|^2} \,, \\[5pt]
\hat{C} &\equiv \left(1+|k|^2\right)^{-2} C = - \frac{1}{\hat{Y}_{22} + Z_2 |H|^2}
\frac{Z_3 \big[ \hat{Y}_{22} + (Z_2+Z_3) |H|^2 \big] - |Z_1|^2 |H|^2 }{ \big[ \hat{Y}_{22} + (Z_2+Z_3) |H|^2 \big]^2 - |Z_1|^2 |H|^4 } \,, \\[8pt]
\Rhat &\equiv \left(1+|k|^2\right)^{1/2} \R = k D^2 H + \hat{J}_2 \,, \\[10pt]
\hat\Phi_{2,\textbf{c}}^{(1)} &\equiv \left(1+|k|^2\right)^{-1/2} \Phi_{2,\textbf{c}}^{(1)}
= - \left[ \hat{A} \big( H^\dagger\Rhat \big)^* + \hat{C} \big( H^\dagger\Rhat \big) \right] H - \hat{B} \Rhat \,.
\end{align}
\end{subequations}
We see that when restricted to two-derivative/fermion order, \ie, the first line of \cref{eq:relevantTermsInEFT}, the matching result is the Standard Model as expected.
Therefore, corrections to the pseudo-observables come from terms at the four- and six-derivative/fermion orders presented in the second line of \cref{eq:relevantTermsInEFT}. To compute these corrections, we will take \cref{eq:relevantTermsInEFT} and expand around the physical vacuum where $H$ has a non-zero vev. We will only keep terms that are relevant for $\kappa_V$ (to six-derivative/fermion order), $\kappa_f$ (to potentially six-derivative/fermion order), and $\kappa_\lambda$ (to four-derivative/fermion order). We also need the propagator residue factors for all the external legs of these amplitudes. It is clear that the four- and six-derivative/fermion terms in \cref{eq:relevantTermsInEFT} do not yield nontrivial corrections to the propagator residues of the gauge bosons or the fermions, but they do modify the Higgs propagator residue factor $Z_h$.

In summary, when we expand \cref{eq:relevantTermsInEFT}, we would like to keep all the terms of the forms
\begin{equation}
h^2\, \partial^n \,,\qquad
h^3\, \partial^n \,,\qquad
W_\mu^+ W_\nu^- h\, \partial^n \,,\qquad
\hat{j}_i\, h\, \partial^n \,,
\label{eqn:terms}
\end{equation}
where $\partial^n$ denotes an arbitrary power of derivatives (up to our truncation order) and $\hat{j}_i$ are the neutral components of $\hat{J}_i$,
\begin{equation}
\hat{J}_i \supset \mqty( 0 \\ \s\hat j_i ) \,.
\end{equation}
Note that all the four- and six-derivative/fermion terms in \cref{eq:relevantTermsInEFT} are quadratic in $\Rhat$. For finding the terms listed in \cref{eqn:terms}, it is therefore sufficient to keep only part of $\Rhat\,$,
\begin{equation}
\Rhat = k D^2 H + \hat{J}_2 \supset \mqty( 0 \\ \frac{k}{\sqrt{2}} \left[ \left( \partial^2 h \right) - \frac12\, g_2^2 v\, W_\mu^+ W^{-\mu} \right] + \hat{j}_2 ) \,,
\end{equation}
and make the replacement
\begin{equation}
H \to \frac{1}{\sqrt{2}} \mqty( 0 \\ v+h ) \,,
\end{equation}
for all the other factors of $H$ fields in the four- and six-derivative/fermion terms in \cref{eq:relevantTermsInEFT} (including the implicit ones in $\hat{A}, \hat{B}, \hat{C}$).
Performing these substitutions, we obtain
\begin{align}
\hspace{-5pt}
\lag_\text{EFT} &\supset \frac12\s (\partial h)^2 - \frac12\s m^2 h^2 - \frac{m^2}{2v}\, h^3
+ \frac12\, g_2^2 v\, W_\mu^+ W^{-\mu} h  -\frac{1}{\sqrt{2}} \frac{v+h}{1+\abs{k}^2}\, \big( \hat{j}_1 + k^* \hat{j}_2 + \text{h.c.} \big)
\notag\\[5pt]
&\hspace{10pt}
+ b_4\, \frac12\, ( \partial^2 h ) \Big[ (\partial^2 h) - g_2^2 v\, W_\mu^+ W^{-\mu} \Big]
+ \left[ \frac{f_4}{\sqrt{2}k^*}\, \hat{j}_2^*\, (\partial^2 h) + \text{h.c.} \right]
+ \frac{\lambda_4}{2}\, h \left(\partial^2 h\right)^2
\notag\\[5pt]
&\hspace{10pt}
+ b_6\, \frac12\, ( \partial_\mu \partial^2 h ) \Big[ (\partial^\mu \partial^2 h) - g_2^2 v\, \partial^\mu(W_\nu^+ W^{-\nu}) \Big]
- \left[ \frac{f_6}{\sqrt{2}k^*}\, \hat{j}_2^*\, (\partial^4 h) + \text{h.c.} \right] \,,
\label{eq:leff}
\end{align}
where the coefficients are
\begin{subequations}
\begin{align}
m^2 &= \lambda_\text{eff}\, v^2 = -2\s m_\text{eff}^2 \,, \\[5pt]
b_4 &= \frac{1}{M_\text{SL}^4} \Bigg\{ \abs{k}^2 \left[ \hat{Y}_{22} + (Z_2+Z_3) \frac{v^2}{2} \right] - \Re\left( k^2 Z_1 \frac{v^2}{2} \right) \Bigg\} \,, \\[5pt]
\Re f_4 &= b_4 \,, \\[5pt]
\Im f_4 &= \frac{1}{M_\text{SL}^4} \Im \left( k^2 Z_1 \frac{v^2}{2} \right) \,, \\[5pt]
\lambda_4 &= \frac{\partial}{\partial v}\s b_4 \,, \\[5pt]
b_6 &= \frac{1}{M_\text{SL}^8} \left( 1 + \abs{k}^2 \right) \abs{ k \left[ \hat{Y}_{22} + (Z_2+Z_3) \frac{v^2}{2} \right] - k^* Z_1^* \frac{v^2}{2} }^2 \,, \\[5pt]
\Re f_6 &= b_6 \,, \\[5pt]
\Im f_6 &= \frac{1}{M_\text{SL}^8} \left( 1 + \abs{k}^2 \right) \left[ \hat{Y}_{22} + (Z_2+Z_3) \frac{v^2}{2} \right] \Im \left( k^2 Z_1 v^2 \right) \,.
\end{align}
\end{subequations}
Note the appearance of the mass scale
\begin{equation}
M^4_\text{SL} = \left[ \hat{Y}_{22} + (Z_2 + Z_3) \frac{v^2}{2} \right]^2 - |Z_1|^2\, \frac{v^4}{4} \,,
\label{eq:SLMassScale}
\end{equation}
which is closely related to the determinant of the mass matrix for the heavy Higgs doublet. (The only difference is the factor of $(1+|k|^2)$ in $\hat{Y}_{22}$, which comes from canonically normalizing $\Phi_1$ to $H$ using \cref{eqn:normalizing}.) $M_\text{SL}$ includes both the explicit mass parameter $Y_{22}$ and the vev-dependent contributions to the mass through the quartic couplings.

From \cref{eq:leff}, the terms that are quadratic in $h$ with no other fields determine that the dispersion relation for $h$ is
\begin{equation}
-m^2+p^2+b_4\s p^4+b_6\s p^6+ O\big(p^8\big)=0 \,,
\end{equation}
which implies that the pole mass $m_h^2$ can be determined by solving 
\begin{equation}
m^2=m_h^2+b_4m_h^4+b_6m_h^6+ O\big(p^8\big) \,,
\end{equation}
and that the residue is 
\begin{align}
Z_h^{-1} &= \pdv{}{p^2} \left(-m^2+p^2+b_4\s p^4+b_6\s p^6+ O\big(p^8\big)\right)\Big|_{p^2=m_h^2} \notag\\[5pt]
&= 1 + 2\s b_4\s m_h^2 + 3\s b_6\s m_h^4 + O\big(p^6\big) \,.
\end{align}
Note that by including higher order momentum terms in the dispersion relation, we are effectively resumming a class of EFT corrections into the propagator.  This is one of the systematic improvements that is facilitated by working in the SL basis.  Using \cref{eq:leff}, we have
\begin{align}
\kappa_V &= Z_h^{1/2}(1+b_4\s m_h^2+b_6\s m_h^4)+ O\big(m_h^6\big) = 1 - \frac12 m_h^4 \left( b_6 - b_4^2 \right) + O\big(m_h^6\big) \,,
\end{align}
and
\begin{equation}
\kappa_\lambda = 1 - 2\s m_h^2\, \frac{\partial}{\partial v^2} \left(v^2 b_4\right) + O\big(m_h^4\big) \,.
\end{equation}
In particular, we note that the quantity appearing in $\kappa_V$ is non-negative,
\begin{equation}
b_6 - b_4^2 = \frac{1}{M_\text{SL}^8} \left\{ \left| k \left[ \hat{Y}_{22} + (Z_2+Z_3) \frac{v^2}{2} \right] - \frac{v^2}{2} k^* Z_1^* \right|^2 + \left[ \frac{v^2}{2} \Im(k^2 Z_1) \right]^2 \right\} \ge 0 \,,
\end{equation}
which guarantees that the correction $\kappa_V - 1 \le 0$ has the correct sign.

Determining $\kappa_f$ is complicated by the fact that there are different possibilities for the fermion couplings to the two doublets. It is most transparent to write the couplings to fermions in the Higgs basis, for which the neutral components of the currents are
\begin{equation}
J_\hg{1} = \mqty(0 \\ j_\hg{1}) \,,\qquad
J_\hg{2} = \mqty(0 \\ j_\hg{2}) \,.
\end{equation}
Using the mappings given in \cref{app:SLToHiggs}, the SL basis currents are then
\begin{equation}
J_a = \frac{1}{\sqrt{1+\abs{k}^2}}\, \mqty( 1 & -k^* \\ k & 1)_{a \hg{b}}\; J_\hg{b} \,.
\end{equation}
This implies
\begin{subequations}
\begin{align}
\hat{j}_1 &= j_\hg{1} - k^* j_\hg{2} \,, \\[3pt]
\hat{j}_2 &= k\s j_\hg{1} + j_\hg{2} \,.
\end{align}
\end{subequations}
The part of \cref{eq:leff} containing fermions can be expressed in terms of Higgs basis currents as
\begin{align}      
\lag_\text{eff} \supset - \frac{1}{\sqrt{2}} j_\hg{1} \left( v + h - f_4^* \partial^2 h + f_6^* \partial^4 h\right) - \frac{1}{\sqrt{2}} \frac{j_\hg{2}}{k} \left(- f_4^* \partial^2 h + f_6^* \partial^4 h\right)  + \text{h.c.} \,.
\end{align}
We see that matching the fermion masses determines $j_\hg{1}$ and places no constraint on $j_\hg{2}$; this is why it is useful to write the Lagrangian in terms of these quantities. The amplitude for a Higgs to decay to a particular chirality of fermions is then proportional to the matrix element of the unconjugated currents,
\begin{align} 
\mathcal{A}_{h \to \bar{f}_L f_R} &=
- \frac{\left\langle j_\hg{1} \right\rangle}{\sqrt{2}}\, Z_h^{1/2} \left(1+ f_4^* m_h^2 + f_6^* m_h^4\right)
- \frac{\left\langle j_\hg{2} \right\rangle}{k \sqrt{2}}\, Z_h^{1/2} \left(f_4^* m_h^2 + f_6^* m_h^4\right) + O\big(m_h^6\big)
\notag\\[10pt]
&=
- \frac{\left\langle j_\hg{1} \right\rangle}{\sqrt{2}} \left[ 1 - \frac12 \left(b_6-b_4^2\right) m_h^4 \right]
+ i\s \frac{\left\langle j_\hg{1} \right\rangle}{\sqrt{2}} \Big[ m_h^2 \Im f_4 + m_h^4 \big( \Im f_6 - b_4 \Im f_4 \big) \Big]
\notag \\[5pt]
&\hspace{12pt}
- \frac{\left\langle j_\hg{2} \right\rangle}{k \sqrt{2}} \left[ b_4\s m_h^2 + \left(b_6-b_4^2\right) m_h^4 \right]
+ i \frac{\left\langle j_\hg{2} \right\rangle}{k \sqrt{2}} \Big[ m_h^2 \Im f_4 + m_h^4 \big( \Im f_6 - b_4 \Im f_4 \big) \Big]
\notag\\[8pt]
&\hspace{12pt}
+ O\big(m_h^6\big) \,,
\end{align}
where we are using a shorthand $\left\langle j \right\rangle = \left \langle \bar{f}_L f_R | j | 0 \right\rangle$.

To calculate $\kappa_f$, we then need to specify $j_\hg{2}$. There are a wide variety of possibilities with rich phenomenological implications, including conventional choices satisfying the Glashow-Weinberg condition \cite{Glashow:1976nt}. In this work, we consider two specific choices. For both, we require the UV 2HDM potential to be CP-preserving; this means $\Im f_4=\Im f_6=0$. For our first example, we set $j_\hg{2}=0$, such that the fermion currents only couple to the linear combination of Higgses that gets a vev. In this case,
\begin{equation}
\kappa_f = 1 - \frac12\, \big( b_6 - b_4^2 \big) m_h^4 = \kappa_V \,.
\end{equation}
In other words, to this order in the EFT expansion there is simply a universal rescaling of all Higgs couplings for this scenario.
This is the unique choice for which $\kappa_f$ does not receive a contribution at leading order. For our second example, we set $j_\hg{2}=j_\hg{1}$, in which case,
\begin{equation}
\kappa_f = 1 + \frac{b_4}{k}\, m_h^2 \,,
\label{eq:kappaf2}
\end{equation}
where we have truncated to the leading order correction. Note that both of the possibilities we consider automatically ensure that there are no FCNC's at tree level.

%%%%%%%%%%%%%%%%%%%%%%%%%%%%%%%%%%%%%%%%%%%%%%%%%%%%%%%%%%%%%%%%%%%%%%%%%%%%%%%%
\section{Numerical Comparison}
\label{sec:Numerics}
%%%%%%%%%%%%%%%%%%%%%%%%%%%%%%%%%%%%%%%%%%%%%%%%%%%%%%%%%%%%%%%%%%%%%%%%%%%%%%%%

We will now provide the results of a scan in the 2HDM parameter space in order to compare the efficacy of the SL basis EFT with the Higgs basis EFT.  We will provide results for the three pseudo-observables derived in the previous section: the shift in the $h\s W^+ W^-$ coupling $\kappa_V$, the shift in the $h^3$ coupling $\kappa_\lambda$, and the shift in the $h\bar{f} f$ coupling $\kappa_f$. For $\kappa_f$, we consider specifically the case when the Yukawa couplings of both doublets are the same in the Higgs basis; see \cref{eq:kappaf2}.  We will present the results in terms of the fractional error of the EFT prediction as compared to the UV prediction,
\begin{equation}
\delta\kappa_{i,\text{EFT}}\equiv\frac{\kappa_{i,\text{EFT}}-\kappa_{i,\text{UV}}}{\kappa_{i,\text{UV}}-1} \,,
\end{equation}
where $\kappa_{i,\text{UV}}$ use the couplings computed in the full 2HDM; both the UV and the Higgs basis EFT results are taken from \cite{Egana-Ugrinovic:2015vgy}.

To make this comparison, we reduce the general 2HDM down to a four-parameter space of models. We first impose custodial symmetry and work in the resulting Higgs basis for which all parameters are real and $Z_{\hg{1}\hg{2}\hg{1}\hg{2}} = Z_{\hg{1}\hg{2}\hg{2}\hg{1}}$. We then scan over the 4 parameters
\begin{equation}
Y_{\hg{1}\hg{2}} \,,\quad
Y_{\hg{2}\hg{2}} \,,\quad
Z_{\hg{1}\hg{1}\hg{1}\hg{1}} \,,\quad
Z_{\hg{1}\hg{1}\hg{2}\hg{2}} \,.
\label{eq:freeParameters}
\end{equation}
Of the remaining parameters, $Y_{\hg{1}\hg{1}}$ and $Z_{\hg{1}\hg{1}\hg{1}\hg{2}}$ are fixed by the Higgs basis vev conditions \cref{eq:HBVevConditions}; the others we fix to satisfy
\begin{equation}
Z_{\hg{1}\hg{2}\hg{2}\hg{2}} = Z_{\hg{1}\hg{2}\hg{1}\hg{2}} = 0 \,;\qquad 
Z_{\hg{2}\hg{2}\hg{2}\hg{2}} = Z_{\hg{1}\hg{1}\hg{1}\hg{1}} \,,
\end{equation}
for simplicity. Note that it is important that $Y_{\hg{1}\hg{2}} \neq 0$ for the Higgs and SL bases to be distinct.

The four free parameters, \cref{eq:freeParameters}, are scanned in units of $v=246 \text{ GeV}$ via a Markov Chain Monte Carlo  (MCMC) method, which samples from the Gaussian likelihood of approximate current experimental constraints on $m_h$ and $\kappa_V \equiv \sin (\beta - \alpha)$. Here $\alpha$ is the familiar Higgs mixing angle and $\sin(\beta - \alpha) \rightarrow 1$ is known as the {\it alignment limit}. Explicitly, we take
\begin{subequations}
\begin{align}
\frac{m_h^2}{v^2} &= 0.2587 \pm 0.0007 \,, \\[3pt]
\kappa_V &= 1.0 \pm 0.1 \,.
\end{align}
\end{subequations}
The MCMC is seeded on a grid of inert 2HDMs, where
\begin{subequations}
\begin{align}
Y_{\hg{1}\hg{2}} &= 0 \,, \\[5pt]
Y_{\hg{2}\hg{2}} &= m_H^2 (1-f) \,, \\[5pt]
Z_{\hg{1}\hg{1}\hg{1}\hg{1}} &= 0.2587 \,, \\[5pt]
Z_{\hg{1}\hg{1}\hg{2}\hg{2}} &= 2 f \frac{m_H^2}{v^2} \,,
\end{align}
\end{subequations}
where $m_H^2$ is the heavy Higgs mass at the global minimum and $f$ is the fraction of it which comes through the cross quartic interaction $\frac12 Z_{\hg{1}\hg{1}\hg{2}\hg{2}} v^2$. We sample $m_H^2$ and $f$ from the discrete sets
\begin{subequations}
\begin{align}
\frac{m_H^2}{\mathrm{GeV}} &= \left\{ 400, 500, 600, 700, 800 \right\} \,, \\[5pt]
f &= \left\{ 0.1, 0.2, 0.3, 0.4, 0.5, 0.6, 0.7, 0.8, 0.9 \right\} \,.
\end{align}
\end{subequations}
Discarding all models with unbounded potentials, we are left with $\sim 7000$ 2HDM model points in the following analysis.

The performance of the SL basis EFT can be understood primarily by looking at two parameters: the alignment of the 2HDM and the mass scale from the mass matrix of the heavy doublet, $M_\text{SL}$, defined in \cref{eq:SLMassScale}. Recall from \cref{subsec:organizing} that the SL basis EFT is an expansion in powers of derivatives (and fermions). We thus expect the $n^\text{th}$ order corrections to our pseudo-observables to scale as
\begin{equation}
\left(D^2\right)^n \sim m_h^{2n} \sim v^{2n} \,.
\end{equation}
By dimensional analysis, the $n^\text{th}$ order corrections must also scale as some mass scale to the power of $-2n$. From Eqs.~(\ref{eq:allorder}), these powers of mass dimension come from the inverse of the mass matrix for the heavy doublet; the $n^\text{th}$ order corrections to our pseudo-observables thus scale as $M_\text{SL}^{-2n}$. The corrections therefore scale as
\begin{equation}
\text{SL basis power counting} \sim \left(\frac{v}{M_\text{SL}}\right)^{2n} \,,
\label{eq:SLpower}
\end{equation}
and we expect that the SL EFT expansion will provide a good approximation when $M_\text{SL}$ is large.

\begin{figure}[t!]
\centering
\includegraphics[align=c,width=.65\textwidth]{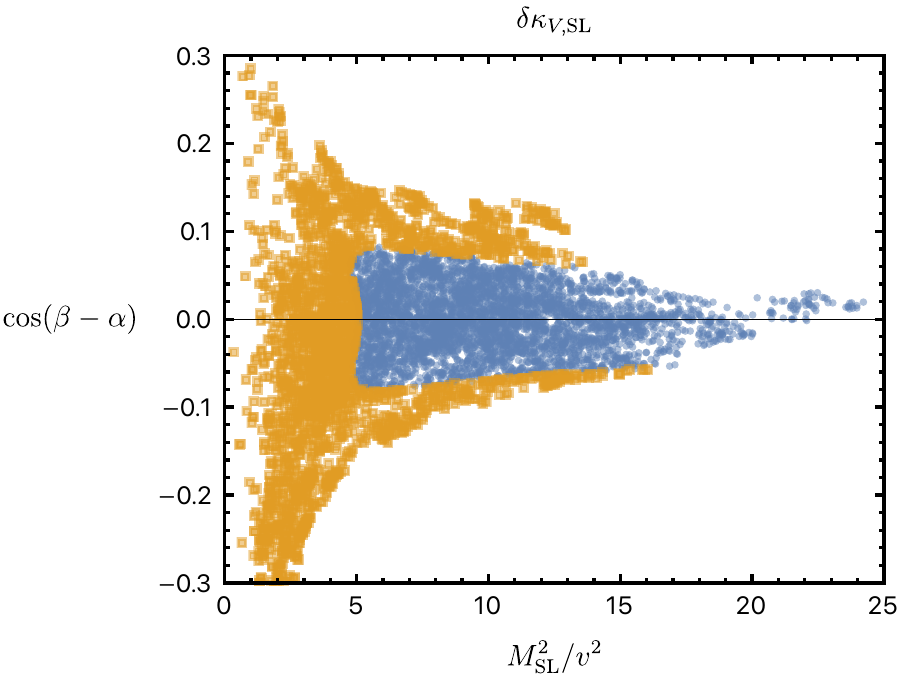}
\caption{This figure shows for which models the SL basis EFT makes an accurate estimate of $\kappa_V$. Blue points are those for which $\delta\kappa_{V,\text{SL}}<0.1$ and orange points (shown on top of the blue points) are those for which $\delta\kappa_{V,\text{SL}}>0.1$.} 
\label{fig:kVSLBad}
\vspace{20pt}
\end{figure}

\begin{figure}[h!]
\centering
\includegraphics[align=c,width=.65\textwidth]{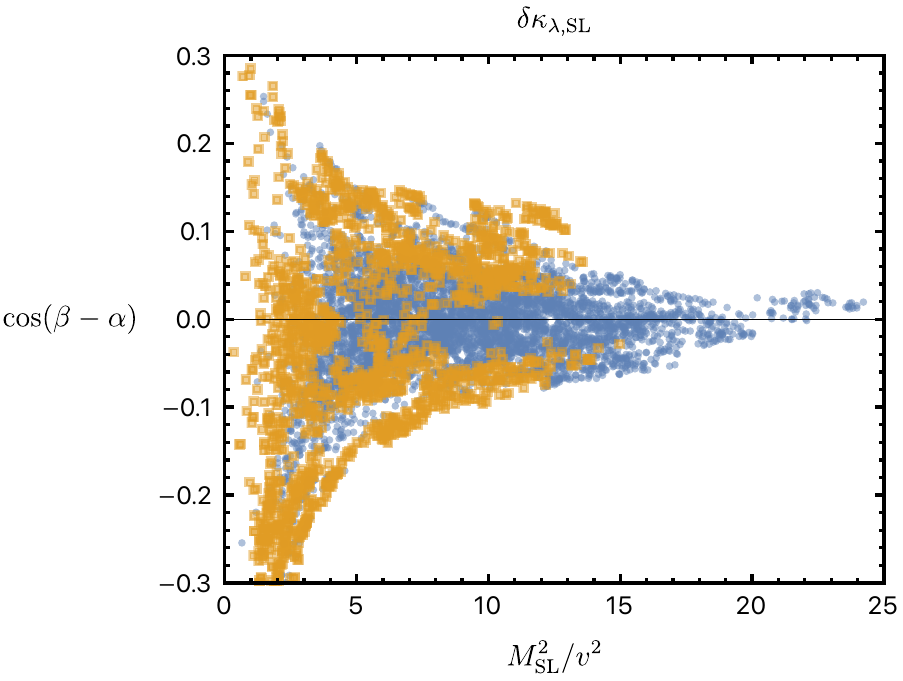}
\caption{This figure shows for which models the SL basis EFT makes an accurate estimate of $\kappa_\lambda$. Blue points are those for which $\delta\kappa_{\lambda,\text{SL}}<0.1$ and orange points (shown on top of the blue points) are those for which $\delta\kappa_{\lambda,\text{SL}}>0.1$.}
\label{fig:klSLBad}
\end{figure}

\begin{figure}[t!]
\centering
\includegraphics[align=c,width=.65\textwidth]{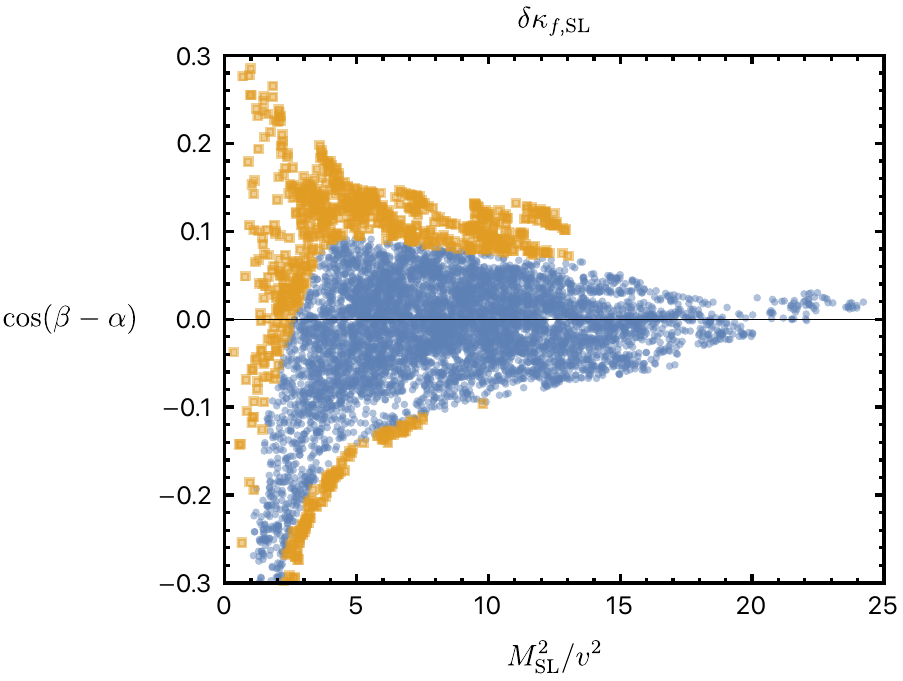}
\caption{This figure shows for which models the SL basis EFT makes an accurate estimate of $\kappa_f$. We have taken the Yukawa couplings to the two Higgs doublets to be equal. Blue points are those for which $\delta\kappa_{f,\text{SL}}<0.1$ and orange points (shown on top of the blue points) are those for which $\delta\kappa_{f,\text{SL}}>0.1$. If the Yukawa couplings of the heavy doublet are instead set to zero, $\kappa_f=\kappa_V$.}
\label{fig:kfSLBad}
\end{figure}

We plot our pseudo-observables in the $\cos(\alpha-\beta)$ versus $M_\text{SL}$ plane in \cref{fig:kVSLBad,fig:kfSLBad,fig:klSLBad}, where $\alpha$ is the Higgs mixing angle, $\beta = \arctan(v_2/v_1)$, and the combination $\cos(\alpha-\beta)$ is a measure of the alignment limit for the 2HDM.  In the figures, we 
separate the points into those for which the fractional error is above or below 10\% to provide a proxy for when the SL basis EFT prediction is accurate.  As expected, we find better performance for larger values of $M_\text{SL}$. In addition, $\kappa_V$ and $\kappa_f$ are highly correlated with the measure of alignment; this is because $\kappa_{V,\text{UV}}$ and $\kappa_{f,\text{UV}}$ depend only on the alignment of the 2HDM (and, for $\kappa_f$, the Yukawa couplings of the Higgs doublets, which we have fixed). For $\kappa_\lambda$, the behavior is more complicated as a larger number of parameters affect the value of $\kappa_\lambda$.

\begin{figure}[t!]
\centering
\includegraphics[align=c,width=.7\textwidth]{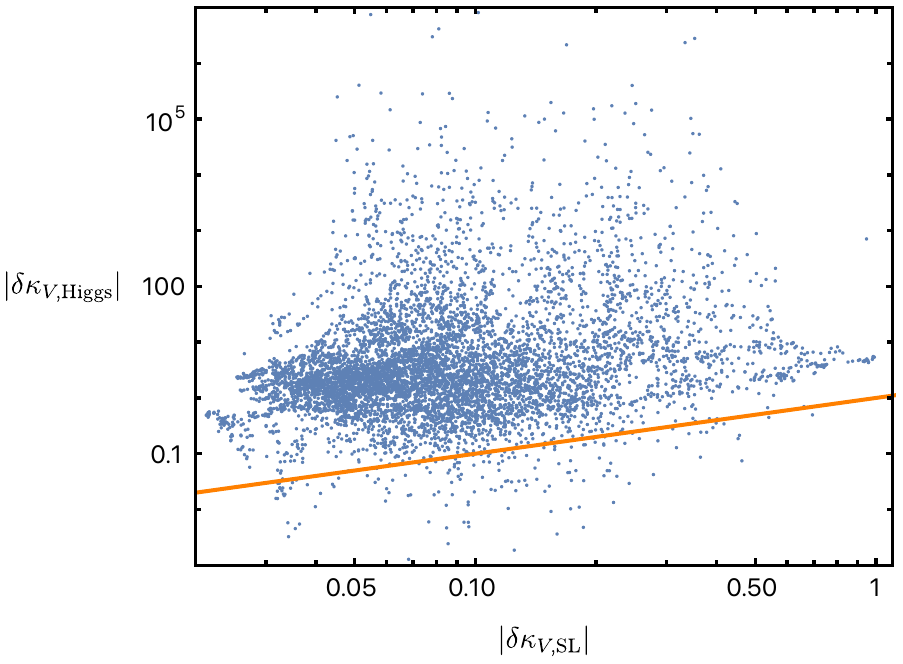}
\caption{This figure shows the accuracy in computing $\kappa_V$ for each of the EFTs. The straight orange line denotes equality between the accuracy of the two EFTs, with points above the line being those for which the SL basis EFT performs better than the Higgs basis EFT.}
\label{fig:kVComp}
\vspace{20pt}
\end{figure}

\begin{figure}[h!]
\centering
\includegraphics[align=c,width=.7\textwidth]{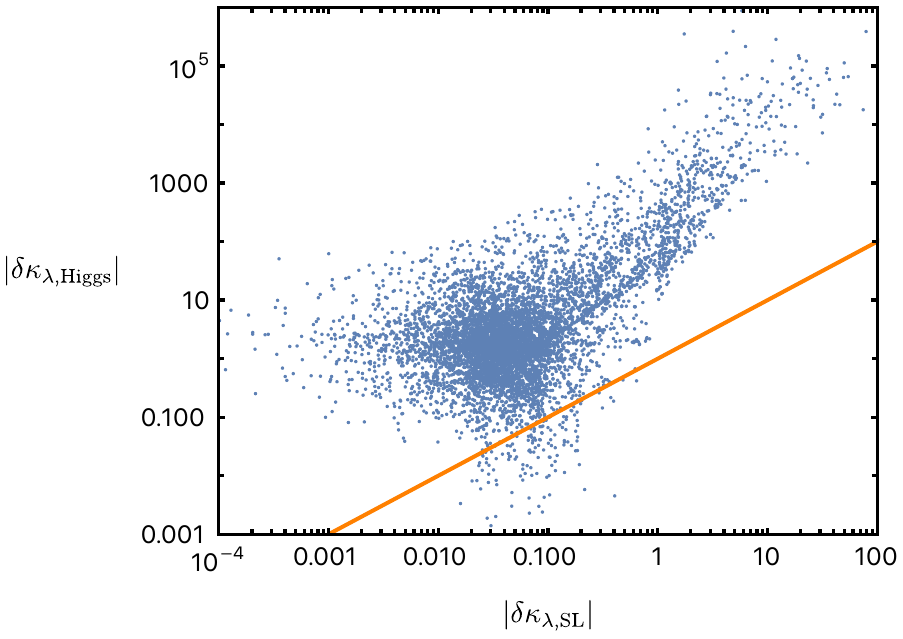}
\caption{This figure shows the accuracy in computing $\kappa_\lambda$ for each of the EFTs. The straight orange line denotes equality between the accuracy of the two EFTs, with points above the line being those for which the SL basis EFT performs better than the Higgs basis EFT.}
\label{fig:klComp}
\end{figure}

\begin{figure}[t!]
\centering
\includegraphics[align=c,width=.7\textwidth]{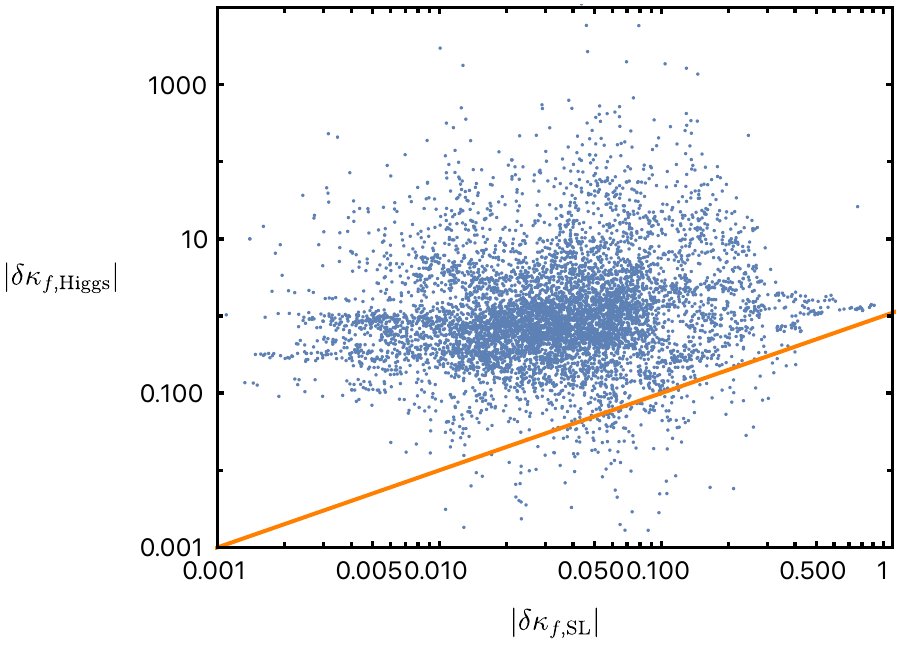}
\caption{This figure shows the accuracy in computing $\kappa_f$. We have taken the Yukawa couplings to the two Higgs doublets to be equal. The straight orange line denotes equality between the accuracy of the two EFTs, with points above the line being those for which the SL basis EFT performs better than the Higgs basis EFT. If the Yukawa couplings of the heavy doublet are instead set to zero, $\kappa_f=\kappa_V$.}
\label{fig:kfComp}
\end{figure}

A comparison of the performance for the SL basis EFT against the Higgs basis EFT is given in \cref{fig:kVComp,fig:kfComp,fig:klComp}. The SL basis EFT typically outperforms the Higgs basis EFT by a significant margin (around 1-2 orders of magnitude smaller fractional error) for all three pseudo-observables; this is the case whether the SL basis EFT performs relatively well or relatively poorly. We do find parameter points for which the Higgs basis EFT outperforms the SL basis EFT, so the SL basis EFT is not universally better. In addition, for a significant minority of points the Higgs basis EFT catastrophically fails with a fractional error of several orders of magnitude; these catastrophic failures include many points on which the SL basis EFT performs quite well. By contrast, while points exist for which the SL basis EFT performs poorly, the Higgs basis tends to perform poorly as well, and none of the points included in our scan show a catastrophic failure of the SL basis.

\begin{figure}[t!]
\centering
\includegraphics[align=c,width=.7\textwidth]{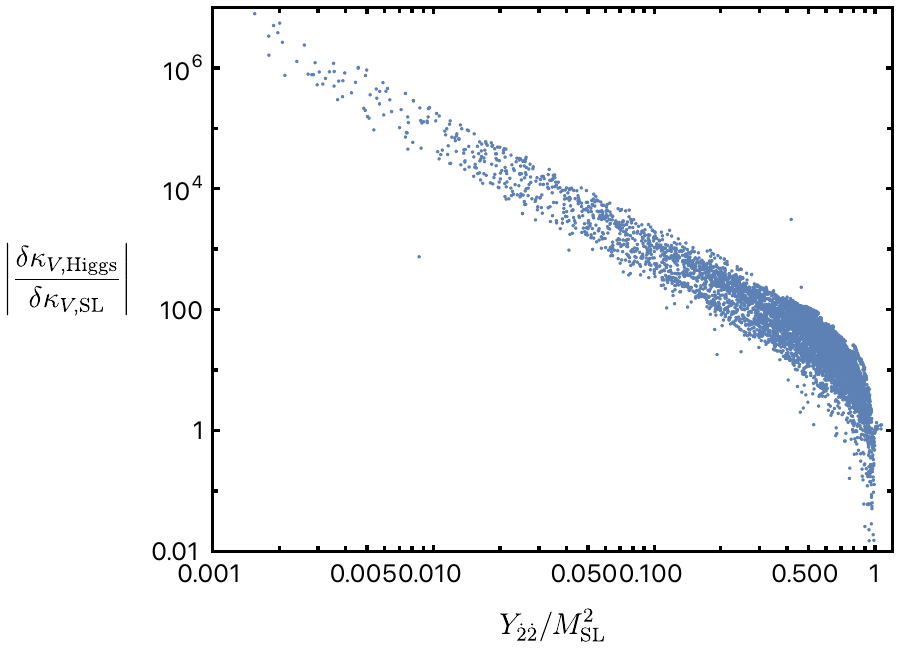}
\caption{This figure shows how the performance of the SL basis EFT and the Higgs basis EFT prediction for $\kappa_V$ depends on the ratio of the mass scales in the two EFTs.}
\label{fig:kVHiggsWorse}
\vspace{40pt}
\end{figure}

\begin{figure}[h!]
\centering
\includegraphics[align=c,width=.7\textwidth]{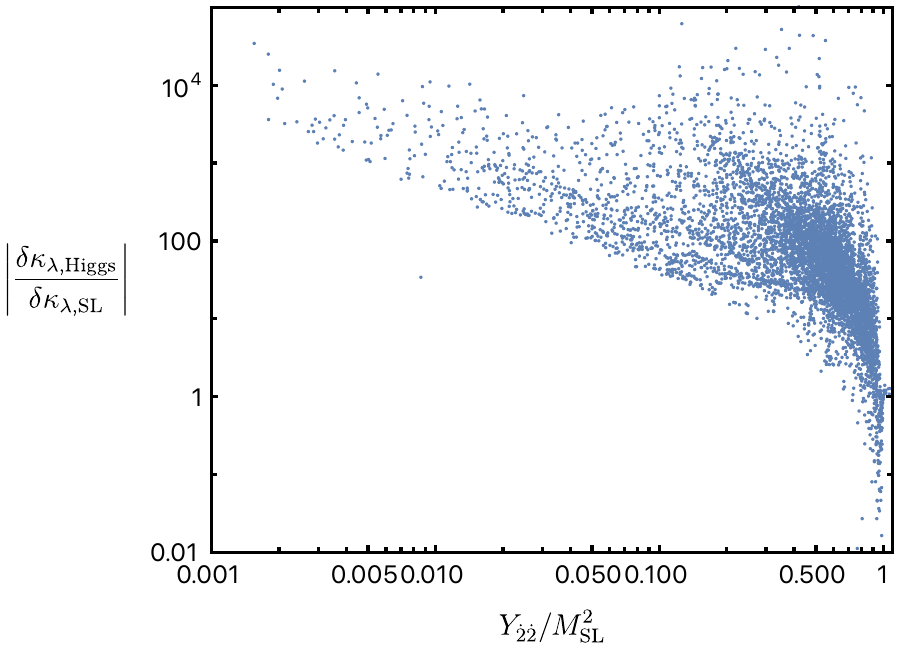}
\caption{This figure shows how the performance of the SL basis EFT and the Higgs basis EFT prediction for $\kappa_\lambda$ depends on the ratio of the mass scales in the two EFTs.}
\label{fig:klHiggsWorse}
\end{figure}

\begin{figure}[t!]
\centering
\includegraphics[align=c,width=.7\textwidth]{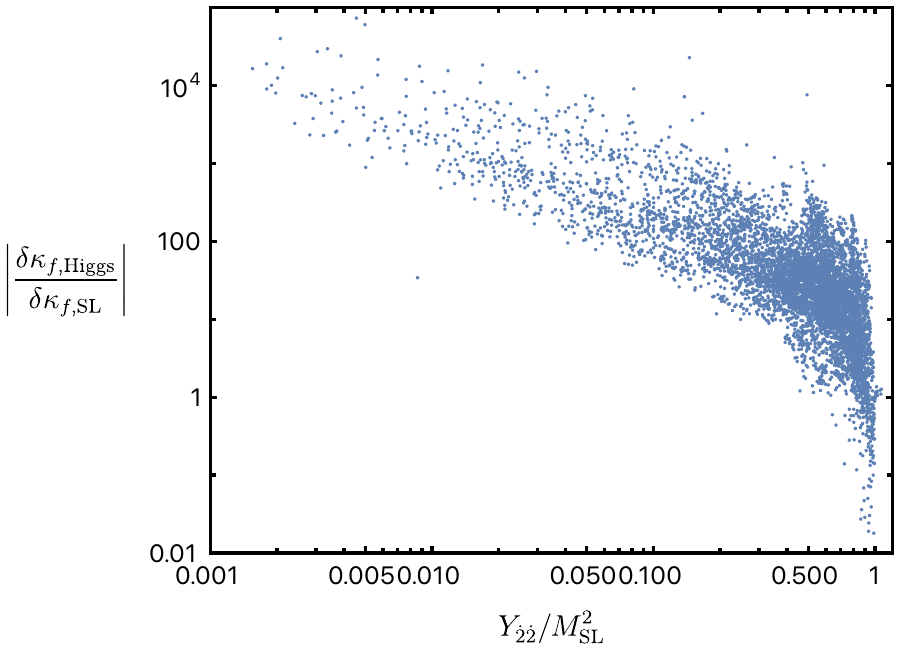}
\caption{This figure shows how the performance of the SL basis EFT and the Higgs basis EFT prediction for $\kappa_f$ depends on the ratio of the mass scales in the two EFTs. We have taken the Yukawa couplings to the two Higgs doublets to be equal. If the Yukawa couplings of the heavy doublet are instead set to zero, $\kappa_f=\kappa_V$.}
\label{fig:kfHiggsWorse}
\end{figure}

The difference in performance of the two EFTs can be understood by looking at the mass scales involved in the power counting. As we saw in \cref{eq:SLpower}, higher-order corrections in the SL basis EFT are suppressed by powers of $M_\text{SL}$; this mass scale is comparable to the physical mass of the second Higgs doublet, so the EFT produces reliable results when the second Higgs doublet is heavy. The higher-order corrections in the Higgs basis EFT are suppressed by powers of $Y_{\hg2\hg2}$. If the second Higgs doublet receives a large contribution to its mass from the vev, then $Y_{\hg2\hg2}$ can be significantly smaller than the mass of the second Higgs. We show how the relative performance of the two EFTs depends on the ratio of their respective mass scales in \cref{fig:kVHiggsWorse,fig:kfHiggsWorse,fig:klHiggsWorse}. We indeed see that when the mass scale of the Higgs basis EFT is significantly smaller than that of the SL basis EFT, the Higgs basis EFT is significantly less accurate. In addition, for all those points on which the Higgs basis EFT is more accurate than the SL basis EFT, the mass scales of the two EFTs are comparable, as expected.

\clearpage

%%%%%%%%%%%%%%%%%%%%%%%%%%%%%%%%%%%%%%%%%%%%%%%%%%%%%%%%%%%%%%%%%%%%%%%%%%%%%%%%
\section{Conclusions}
\label{sec:conclusions}
%%%%%%%%%%%%%%%%%%%%%%%%%%%%%%%%%%%%%%%%%%%%%%%%%%%%%%%%%%%%%%%%%%%%%%%%%%%%%%%%

In this paper, we have derived the tree-level matching coefficients by integrating out the BSM states in the 2HDM.  The novel aspect of this work is the introduction of the SL basis, which is an optimal choice for performing the matching calculation.  Working with the SL basis allows us to match a far broader parameter space of 2HDM models onto SMEFT and to resum all orders of the light Higgs field into the EFT Wilson coefficients in a systematic way.  This leads to significantly improved predictions when compared to the computation performed using the Higgs basis in the UV across most of the 2HDM parameter space.  This demonstrates the utility of the EFT derived using the SL basis.  In particular, this is the basis to use if one is interested in exploring the EFT predictions for the 2HDM for models that have alignment away from the decoupling limit. This brings the 2HDM fully into the EFT fold, extending the validity of EFT interpretations of Higgs coupling measurements across a wider range of 2HDMs.

There are many future directions to explore. As we have worked strictly at tree level, extending the SL basis EFT matching calculation to loop level (potentially with functional matching techniques) is a natural next step. Although we have focused our numerical studies on CP-conserving 2HDM, the SL basis is applicable in the fully general CP-violating 2HDMs, where further numerical studies are likely to be informative. It would also be instructive to extend the SL basis to models with extra scalar fields beyond the 2HDM. 

There are also more phenomenological studies that could be done. Our expressions shed light on the physical combinations of parameters that can appear in the low energy virtual effects of the heavy doublet. However, since we only explored the properties of pseudo-observables here, it would be important to compute a set of full LHC observables which would serve as inputs to provide constraints on the 2HDM parameter space.  It is possible that one could then identify novel indirect searches that could be performed which would be particularly sensitive to the effects of the 2HDM.  And in the event that an indirect signal of BSM physics would be discovered, the results here would facilitate our ability to interpret such a signal in terms of the 2HDM parameter space.

\acknowledgments

We thank Howie Haber and Carlos Wagner for useful conversations and thank Howie Haber for useful comments on the manuscript.
The work of I.~Banta and N.~Craig is supported by the U.S.~Department of Energy under the grant DE-SC0011702.
The work of T.~Cohen is supported by the U.S.~Department of Energy under grant number DE-SC0011640.
X.~Lu is supported by the U.S.~Department of Energy under grant number DE-SC0009919.
D.~Sutherland has received funding from the European Union's Horizon 2020 research and innovation programme under the Marie Skłodowska-Curie grant agreement No.~754496.
The work of T.~Cohen, N.~Craig, X.~Lu, and D.~Sutherland was performed in part at the Aspen Center for Physics, which is supported by National Science Foundation grant PHY-1607611. The participation of D.~Sutherland at the Aspen Center for Physics was supported by the Simons Foundation.

\appendix
\section*{Appendices}
\addcontentsline{toc}{section}{\protect\numberline{}Appendices}%

%%%%%%%%%%%%%%%%%%%%%%%%%%%%%%%%%%%%%%%%%%%%%%%%%%%%%%%%%%%%%%%%%%%%%%%%%%%%%%%%
\section{Mapping EFT Quantities from SL to Higgs Basis}
\label{app:SLToHiggs}
%%%%%%%%%%%%%%%%%%%%%%%%%%%%%%%%%%%%%%%%%%%%%%%%%%%%%%%%%%%%%%%%%%%%%%%%%%%%%%%%

Using \cref{eq:Udef,eq:kAsSLParams,eq:kAsHParams}, quantities appearing in the EFT derived using the SL basis can be written in terms of Higgs basis parameters as
\begin{subequations}
\begin{align}
\hat{Y}_{22} = (1+\abs{k}^2) Y_{22} &= Y_{\hg{2}\hg{2}} + k^* Y_{\hg{2}\hg{1}} + k Y_{\hg{1}\hg{2}} + \abs{k}^2 Y_{\hg{1}\hg{1}}
\notag\\[5pt]
&= Y_{\hg{2}\hg{2}} - \frac{\abs{Y_{\hg{1}\hg{2}}}^2}{Y_{\hg{1}\hg{1}}}  = Y_{\hg{2}\hg{2}} - \abs{k}^2 Y_{\hg{1}\hg{1}} \,, \label{eqn:Y22inHiggs} \\[5pt]
Z_1 &= (1+\abs{k}^2) Z_{\hg{1} 2 \hg{1} 2}
\notag\\[5pt]
&= Z_{\hg{1}\hg{2}\hg{1}\hg{2}} - \frac{Z_{\hg{1}\hg{1}\hg{1}\hg{2}}^2}{Z_{\hg{1}\hg{1}\hg{1}\hg{1}}} = Z_{\hg{1}\hg{2}\hg{1}\hg{2}}  - (k^*)^2 Z_{\hg{1}\hg{1}\hg{1}\hg{1}} \,, \\[5pt]
Z_2 &= (1+\abs{k}^2) Z_{\hg{1} \hg{1} 2 2}
\notag\\[5pt]
&= Z_{\hg{1}\hg{1}\hg{2}\hg{2}} - \frac{\abs{Z_{\hg{1}\hg{1}\hg{1}\hg{2}}}^2}{Z_{\hg{1}\hg{1}\hg{1}\hg{1}}} = Z_{\hg{1}\hg{1}\hg{2}\hg{2}} - \abs{k}^2 Z_{\hg{1}\hg{1}\hg{1}\hg{1}} \,, \\[5pt]
Z_3 &= (1+\abs{k}^2) Z_{\hg{1} 2 2 \hg{1}}
\notag\\[5pt]
&= Z_{\hg{1}\hg{2}\hg{2}\hg{1}}  - \frac{\abs{Z_{\hg{1}\hg{1}\hg{1}\hg{2}}}^2}{Z_{\hg{1}\hg{1}\hg{1}\hg{1}}} = Z_{\hg{1}\hg{2}\hg{2}\hg{1}}  - \abs{k}^2 Z_{\hg{1}\hg{1}\hg{1}\hg{1}} \,, \\[5pt]
Z_4 &= \sqrt{1+\abs{k}^2} Z_{\hg{1} 2 2 2}
\notag\\[5pt]
&= \frac{ 2 \abs{k}^2 Z_{\hg{1}\hg{1}\hg{1}\hg{2}} + k^* Z_{\hg{1}\hg{2}\hg{2}\hg{1}} + k^* Z_{\hg{1}\hg{1}\hg{2}\hg{2}} + k Z_{\hg{1}\hg{2}\hg{1}\hg{2}} + Z_{\hg{1}\hg{2}\hg{2}\hg{2}} }{1+\abs{k}^2}  \,,
\\[5pt]
m^2_\text{eff} &= Y_{\hg{1}\hg{1}} \,, \label{eqn:meffinHiggs} \\[5pt]
\lambda_\text{eff} &= Z_{\hg{1}\hg{1}\hg{1}\hg{1}} \,.
\end{align}
\end{subequations}
In terms of the more conventional 2HDM parameters in Higgs basis (see \cref{eq:mapToMsAndLambdas}), the map is given by
\begin{subequations} \label{eq:SLBasisToMsAndLambdas}
\begin{align}
k &= -\frac{\lambda_6^*}{\lambda_1} \,, \\[5pt]
\hat{Y}_{22} &= m_2^2 - \abs{k}^2 m_1^2 \,, \label{eq:hatY} \\[5pt]
Z_1 &= \lambda_5 - (k^*)^2 \lambda_1 \,, \\[5pt]
Z_2 &= \lambda_3 - \abs{k}^2 \lambda_1  \,, \\[5pt]
Z_3 &= \lambda_4 - \abs{k}^2 \lambda_1 \,, \\[5pt]
Z_4 &= \frac{2 \abs{k}^2 \lambda_6 + k^* \lambda_3 + k^* \lambda_4 + k \lambda_5 + \lambda_7 }{1+\abs{k}^2} \\[5pt]
m^2_\text{eff} &= m_1^2 \,, \\[5pt]
\lambda_\text{eff} &= \lambda_1 \,.
\end{align}
\end{subequations}
Armed with these expressions, we can expand our expressions for our pseudo-observables in the SL basis EFT and check that they agree with the Higgs basis EFT. The Higgs basis EFT is an expansion in inverse powers of $m_2^2$; from \cref{eq:hatY}, we see that this is equivalent to an expansion in inverse powers of $\hat Y_{22}$ (to leading order), so we should expand our expressions for $\kappa_V, \kappa_\lambda, \kappa_f$ in the SL basis EFT to leading order in $\hat Y_{22}$ and then convert to Higgs basis quantities. We have
\begin{align}
b_4 &\simeq \frac{1}{\hat Y_{22}}\, |k|^2 \,, \\[5pt]
b_6 &\simeq \frac{1}{\hat Y_{22}^2}\, |k|^2 (1+|k|^2) \,, \\[5pt]
\lambda_4 &\simeq -v\, \frac{1}{\hat Y_{22}^2}\, \Big[ |k|^2 (Z_2+Z_3) + \Re \left( k^2 Z_1 \right) \Big] \,,
\end{align}
which gives
\begin{align}
\kappa_{V,\text{SL}} &= 1 - \frac12 \left(b_6-b_4^2\right)\, m_h^4 \simeq 1 - \frac12 \frac{|k|^2 \lambda_\text{eff}^2\, v^4}{\hat Y_{22}^2} \simeq 1 - \frac12 \frac{|\lambda_6|^2v^4}{m_2^4} \,, \\[5pt]
\kappa_{f,\text{SL}} &= 1 + \frac{b_4}{k}\, m_h^2 \simeq 1 + \frac{k^*}{\hat Y_{22}}\, \lambda_\text{eff}\, v^2 \simeq  1 - \frac{\lambda_6 v^2}{m_2^2} \,, \\[5pt]
\kappa_{\lambda,\text{SL}} &= 1 - 2\s b_4\s m_h^2 - \lambda_4\s v\s m_h^2 \simeq 1 - 2\s \frac{|k|^2\lambda_\text{eff}\, v^2}{\hat Y_{22}} = 1 - 2\s \frac{|k|^2\lambda_\text{eff}^2\, v^4}{\hat Y_{22} m_h^2} \simeq 1 - 2\s \frac{|\lambda_6|^2 v^4}{m_2^2 m_h^2} \,,
\end{align}
all of which agree with the corresponding Higgs basis EFT expressions in Ref.~\cite{Egana-Ugrinovic:2015vgy}.

%%%%%%%%%%%%%%%%%%%%%%%%%%%%%%%%%%%%%%%%%%%%%%%%%%%%%%%%%%%%%%%%%%%%%%%%%%%%%%%%
\section{Equivalence of Decoupling-Limit SL and Higgs Basis EFTs}
\label{app:Converting}
%%%%%%%%%%%%%%%%%%%%%%%%%%%%%%%%%%%%%%%%%%%%%%%%%%%%%%%%%%%%%%%%%%%%%%%%%%%%%%%%

The SL basis and Higgs basis EFTs are generally two different EFTs, equipped with their own power counting, and regimes of validity. However, in the decoupling limit, we can expand in inverse powers of large $Y_{\hg{2}\hg{2}}\equiv m_2^2$ to reproduce the same effects.

At the Lagrangian level this manifests as a field redefinition equivalence between the two EFTs. Whereas the two bases are related by a simple non-derivative field redefinition in the UV, in the EFT this requires a more complicated field redefinition with derivatives \cite{Criado:2018sdb}. Here, we show the equivalence explicitly within the scalar parts of the two EFTs, working in the custodial limit and up to dimension 8 order, \ie, $O(1/m_2^4)$.

Expanding the SL Basis EFT \cref{eq:relevantTermsInEFT} in the custodial limit, we find
\begin{align}
\lag =& - H^\dagger D^2 H - m_1^2 \abs{H}^2 - \frac12 \lambda_1 \abs{H}^4 + \left(\frac{k^2}{m_2^2} + \frac{k^4 m_1^2}{m_2^4}\right) \abs{D^2 H}^2 - \frac{k^2 Z_2}{m_2^4} \abs{H}^2 \abs{D^2 H}^2
\notag\\[5pt]
& - \frac{k^2 Z_1}{2 m_2^4} \left( H^\dagger D^2 H + \text{h.c.} \right)^2 - \frac{k^2 (1+k^2)}{m_2^4} \left(D^2 H^\dagger\right) \left(D^4 H\right) \,.
\label{eq:SLBasisEFTUntransformed}
\end{align}
We have used \cref{eq:SLBasisToMsAndLambdas} to convert the Wilson coefficients to Higgs basis parameters. Under the substitution
\begin{align}
H &\to H + \left\{ -\frac{k^2 \lambda_1}{2 m_2^2} + \frac{k^2 m_1^2}{4 m_2^4} \Big[ 4 Z_1 + 2 Z_2 - ( 4 - k^2) \lambda_1 \Big] \right\} \abs{H}^2 H
\notag\\[5pt]
&\hspace{15pt}
+ \frac{k^2 \lambda_1}{8 m_2^4} \Big[ 8 Z_1 + 4 Z_2 - (4 - 9 k^2) \lambda_1 \Big] \abs{H}^4 H
\notag\\[5pt]
&\hspace{15pt}
+ \left[ \frac{k^2}{2 m_2^2} + \frac{ k^2 (1 + k^2) m_1^2}{2 m_2^4} \right] (D^2 H)
- \frac{k^2}{4 m_2^4} \Big[ 2 Z_2 - (2 - k^2) \lambda_1 \Big] \abs{H}^2 (D^2 H)
\notag\\[5pt]
&\hspace{15pt}
+ \frac{k^2 (4 Z_1 + k^2 \lambda_1)}{4 m_2^4} \abs{D_\mu H}^2 H
- \frac{k^2}{4 m_2^4} \Big[ 2 Z_1 - (2 - k^2) \lambda_1 \Big] \left(D^2 \abs{H}^2\right) H
\notag\\[5pt]
&\hspace{15pt}
+ \frac{k^2 \lambda_1 (4 - k^2)}{4 m_2^4} \left(D_\mu \abs{H}^2\right) \left(D^\mu H\right) 
-\frac{k^2 (4 + k^2)}{8 m_2^4} (D^4 H) \,,
\end{align}
and with the use of the identity
\begin{equation}
2\s \abs{D_\mu H}^2 = D^2 \abs{H}^2 - \left( H^\dagger D^2 H + \text{h.c.} \right) \,,
\end{equation}
and the integration-by-parts relations
\begin{subequations}
\begin{align}
-2\, \abs{H}^2 \left(D_\mu \abs{H}^2\right) \left(D^\mu \abs{H}^2\right) &= \abs{H}^4 \left( D^2 \abs{H}^2 \right) \,, \\[10pt]
2\, \abs{H}^2 H^\dagger D^2 \left( \abs{H}^2 H \right) &= \abs{H}^4 \left( H^\dagger D^2 H + \text{h.c.}\right) + \abs{H}^4 \left( D^2 \abs{H}^2 \right) \,, \\[8pt]
\left( D_\mu \abs{H}^2 \right) \left[ \left( D^\mu H \right)^\dagger \left(D^2 H\right) + \text{h.c.} \right] &= - 2 \abs{H}^2 \abs{D^2 H}^2
\notag\\[2pt]
&\quad
- \abs{H}^2 \left[ \left(D_\mu H\right)^\dagger \left( D^\mu D^2 H \right) + \text{h.c.}\right] \,, \\[8pt]
\left( D^2 \abs{H}^2 \right) \left( H^\dagger D^2 H + \text{h.c.} \right) &= 2 \abs{H}^2 \abs{D^2 H}^2
\notag\\[3pt]
&\quad
+ 2 \abs{H}^2 \left[ \left(D_\mu H\right)^\dagger \left( D^\mu D^2 H \right) + \text{h.c.}\right]
\notag\\[3pt] 
&\quad
+ \abs{H}^2 \left( H^\dagger D^4 H + \text{h.c.} \right) \,,
\end{align}
\end{subequations}
together with the subsequent rescaling
\begin{equation}
H \to H \left[ 1 - \frac{k^2 m_1^2}{2 m_2^2} - \frac{k^2 m_1^4 (4+k^2)}{8 m_2^4} \right] \,,
\end{equation}
the expanded SL Basis EFT in \cref{eq:SLBasisEFTUntransformed} can be reduced to
\begin{align}
\lag &= - H^\dagger D^2 H - \frac{k^2 \lambda_1^2}{2 m_2^4}\, \abs{H}^4 \left( D^2 \abs{H}^2 \right)
- \left( m_1^2 - k^2 \frac{m_1^4}{m_2^2} - k^2 \frac{m_1^6}{m_2^4}\right) \abs{H}^2
\notag \\[5pt]
&\quad
- \frac12 \left[ \lambda_1 -4 k^2 \lambda_1 \frac{m_1^2}{m_2^2}+2 k^2 \frac{m_1^4}{m_2^4} \left( 2 \lambda_4 + \lambda_3 - 3 \lambda_1 \right) \right] \abs{H}^4
\notag\\[5pt]
&\quad
+ \left[ k^2 \frac{\lambda_1^2}{m_2^2}-k^2 \lambda_1 \frac{m_1^2}{m_2^4} \left( 4 \lambda_4 + 2 \lambda_3 - 3 \lambda_1 \right) \right] \abs{H}^6
- k^2 \frac{\lambda_1^2}{m_2^4} \left( 2 \lambda_4 + \lambda_3 - \lambda_1 \right) \abs{H}^8 \,.
\label{eq:fieldRedefinedSLLag}
\end{align}
We have used \cref{eq:SLBasisToMsAndLambdas} again to write $Z_1 = Z_3 = \lambda_4 - k^2 \lambda_1$ and $Z_2 = \lambda_3 - k^2 \lambda_1$ in the custodial limit.

We can compare \cref{eq:fieldRedefinedSLLag} to known results in the Higgs basis EFT. Assuming custodial symmetry, the scalar sector of the general results in \cite{Egana-Ugrinovic:2015vgy} reduce to
\begin{align}
\lag &= \left( 1 + \frac{m_{12}^4}{m_2^4} \right) |D_\mu H|^2 + \frac{2 m_{12}^2 \lambda_6}{m_2^4} \left[ \frac{1}{2} \left(\partial_\mu |H|^2\right) \left(\partial^\mu |H|^2\right) + |H|^2 |D_\mu H|^2 \right]
\notag\\[5pt]
&\quad
+ \frac{\lambda_6^2}{m_2^4} \Big[ 2\, |H|^2 \left(\partial_\mu |H|^2\right) \left(\partial^\mu |H|^2\right) + |H|^4 |D_\mu H|^2 \Big]
\notag\\[5pt]
&\quad
- \left( m_1^2 -  \frac{m_{12}^4}{m_2^2} \right) \abs{H}^2
- \frac12 \left[ \lambda_1 -\frac{4 m_{12}^2 \lambda_6}{m_2^2} + \frac{ 2 (\lambda_3 + 2 \lambda_4) m_{12}^4}{m_2^4} \right] \abs{H}^4
\notag\\[5pt]
&\quad
+ \left[ \frac{\lambda_6^2}{m_2^2} - \frac{2(\lambda_3 + 2 \lambda_4) m_{12}^2 \lambda_6}{m_2^4} \right] |H|^6 
-\frac{ (\lambda_3 + 2 \lambda_4) \lambda_6^2}{m_2^4} \abs{H}^8 \,.
\end{align}
This can be canonically normalized to $O\left( 1/m_2^4 \right)$ to give
\begin{align}
\lag &= |D_\mu H|^2 + \frac{2 m_{12}^2 \lambda_6}{m_2^4} \left[ \frac{1}{2} \left(\partial_\mu |H|^2\right) \left(\partial^\mu |H|^2\right) + |H|^2 |D_\mu H|^2 \right]
\notag\\[5pt]
&\quad
+ \frac{\lambda_6^2}{m_2^4} \Big[ 2\, |H|^2 \left(\partial_\mu |H|^2\right) \left(\partial^\mu |H|^2\right) + |H|^4 |D_\mu H|^2 \Big]
\notag\\[5pt]
&\quad
- \left( m_1^2 - \frac{m_{12}^4}{m_2^2} -\frac{m_{12}^4 m_1^2}{m_2^4} \right) \abs{H}^2
- \frac12 \left[ \lambda_1 - \frac{4 m_{12}^2 \lambda_6}{m_2^2} + \frac{ 2 (\lambda_3 + 2 \lambda_4 - \lambda_1) m_{12}^4}{m_2^4} \right] \abs{H}^4
\notag\\[5pt]
&\quad
+ \left[ \frac{\lambda_6^2}{m_2^2} - \frac{2(\lambda_3 + 2 \lambda_4) m_{12}^2 \lambda_6}{m_2^4} \right] |H|^6
- \frac{ (\lambda_3 + 2 \lambda_4) \lambda_6^2}{m_2^4} \abs{H}^8 \,.
\end{align}
Using IBPs and the field redefinition
\begin{equation}
H \to H - \frac{m_{12}^2 \lambda_6}{m_2^4}\, \abs{H}^2 H - \frac{\lambda_6^2}{2 m_2^4}\, \abs{H}^4 H \,,
\end{equation}
we obtain
\begin{align}
\lag &= |D_\mu H|^2 - \frac{\lambda_6^2}{2 m_2^4} |H|^4 \left( D^2 |H|^2 \right)
- \left( m_1^2 - \frac{m_{12}^4}{m_2^2} -\frac{m_{12}^4 m_1^2}{m_2^4} \right) \abs{H}^2
\notag\\[5pt]
&\quad
- \frac12 \left[ \lambda_1 - \frac{4 m_{12}^2 \lambda_6}{m_2^2} + \frac{ 2 (\lambda_3 + 2 \lambda_4 - 3 \lambda_1) m_{12}^4}{m_2^4} \right] \abs{H}^4
\notag\\[5pt]
&\quad
+ \left[ \frac{\lambda_6^2}{m_2^2} - \frac{ ( 2 \lambda_3 + 4 \lambda_4 - 3 \lambda_1 ) m_{12}^2 \lambda_6}{m_2^4}  \right] |H|^6
- \frac{ (\lambda_3 + 2 \lambda_4-\lambda_1) \lambda_6^2}{m_2^4} \abs{H}^8 \,,
\label{eq:fieldRedefinedHLag}
\end{align}
where we have used $m_1^2 \lambda_6 = m_{12}^2 \lambda_1$ (a consequence of the vev condition in the Higgs basis). As both $k \lambda_1 = -\lambda_6$ and $k\s m_1^2 = -m_{12}^2$ (in the custodial limit), we see \cref{eq:fieldRedefinedHLag} and \cref{eq:fieldRedefinedSLLag} are equivalent.

\addcontentsline{toc}{section}{\protect\numberline{}References}%
\bibliographystyle{JHEP}
\bibliography{2HDM}

\end{document}